\documentclass{aastex63}
\pdfoutput=1 
\usepackage{appendix}
\usepackage{amsmath,amstext}
\usepackage[figure,figure*]{hypcap}
\usepackage{changepage} 
\usepackage{rotating}
\usepackage{scrextend}
\usepackage{lineno}

\providecommand{\msun}{\ensuremath{\,M_{\odot}}}

\providecommand{\mj}{\ensuremath{\,M_{\rm J}}}

\providecommand{\me}{\ensuremath{\,M_{\rm \oplus}}}

\newcommand{\tess}{\emph{TESS}}
\let\orgautoref\autoref
\renewcommand{\autoref}
    {\def\equationautorefname{Eq.}%
     \def\figureautorefname{Fig.}%
     \def\sectionautorefname{Sect.}%
     \def\subsectionautorefname{Sect.}%
     \def\subsubsectionautorefname{Sect.}%
     \orgautoref}
\newcommand{\mos}{\,m\,s$^{-1}$}
\newcommand{\kms}{\,km\,s$^{-1}$}

\renewcommand{\deg}{^{\circ}}
\submitjournal{Astronomical Journal}

\shorttitle{HD 83443c}
\shortauthors{Errico et al.}

\graphicspath{{./}{figures/}}

\begin{document}


\title{HD 83443c: A highly eccentric giant planet on a 22-year orbit}

\correspondingauthor{Adriana Errico}
\email{Adriana.Errico@usq.edu.au}

\author[0000-0002-0839-4257]{Adriana Errico}
\affiliation{University of Southern Queensland, Centre for Astrophysics, West Street, Toowoomba, QLD 4350 Australia}


\author[0000-0001-9957-9304]{Robert A. Wittenmyer}
\affiliation{University of Southern Queensland, Centre for Astrophysics, West Street, Toowoomba, QLD 4350 Australia}

\author[0000-0002-1160-7970]{Jonathan Horner}
\affiliation{University of Southern Queensland, Centre for Astrophysics, West Street, Toowoomba, QLD 4350 Australia}

\author[0000-0002-4860-7667]{Zhexing Li}
\affil{Department of Earth and Planetary Sciences, University of California, Riverside, CA 92521, USA}

\author[0000-0003-0168-3010]{G.~Mirek Brandt}
\altaffiliation{NSF Graduate Research Fellow}
\affiliation{Department of Physics, University of California, Santa Barbara, Santa Barbara, CA 93106, USA}

\author[0000-0002-7084-0529]{Stephen R. Kane}
\affil{Department of Earth and Planetary Sciences, University of California, Riverside, CA 92521, USA}

\author[0000-0002-3551-279X]{Tara Fetherolf}
\altaffiliation{UC Chancellor's Fellow}
\affil{Department of Earth and Planetary Sciences, University of California, Riverside, CA 92521, USA}

\author[0000-0003-0437-3296]{Timothy R. Holt}
\affil{University of Southern Queensland, Centre for Astrophysics, West Street, Toowoomba, QLD 4350 Australia}
\affil{Department of Space Studies, Southwest Research Institute, Boulder, CO. USA}

\author[0000-0003-0035-8769]{Brad Carter}
\affiliation{University of Southern Queensland, Centre for Astrophysics, West Street, Toowoomba, QLD 4350 Australia}

\author[0000-0003-3964-4658]{Jake T. Clark}
\affiliation{University of Southern Queensland, Centre for Astrophysics, West Street, Toowoomba, QLD 4350 Australia}


\author[0000-0003-1305-3761]{R.P. Butler}
\affiliation{Department of Terrestrial Magnetism, Carnegie Institution of Washington, 5241 Broad Branch Road, NW, Washington, DC 20015-1305, USA}

\author[0000-0002-7595-0970]{C.G. Tinney}
\affil{Exoplanetary Science at UNSW, School of Physics, UNSW Sydney, NSW, 2052, Australia}

\author[0000-0002-3247-5081]{Sarah Ballard}
\affiliation{Department of Astronomy, University of Florida, 211 Bryant Space Science Center, Gainesville, FL, 32611, USA}

\author[0000-0003-2649-2288]{Brendan P. Bowler}
\affil{Department of Astronomy, The University of Texas at Austin, TX 78712, USA}

\author[0000-0003-0497-2651]{John Kielkopf}
\affil{Department of Physics and Astronomy, University of Louisville, Louisville, KY 40292, USA}

\author{Huigen Liu}
\affil{School of Astronomy and Space Science, Key Laboratory of Modern Astronomy and Astrophysics in Ministry of Education, Nanjing University, Nanjing 210046, Jiangsu, China}

\author[0000-0002-8864-1667]{Peter P. Plavchan}
\affiliation{Department of Physics and Astronomy, George Mason University, 4400 University Drive, Fairfax, VA 22030, USA}

\author[0000-0002-1836-3120]{Avi Shporer}
\affil{Department of Physics and Kavli Institute for Astrophysics and Space Research, Massachusetts Institute of Technology, Cambridge, MA 02139, USA}

\author[0000-0003-3491-6394]{Hui Zhang}
\affil{Shanghai Astronomical Observatory, Chinese Academy of Sciences, Shanghai 200030, China}

\author[0000-0001-7294-5386]{Duncan J. Wright}
\affil{University of Southern Queensland, Centre for Astrophysics, West Street, Toowoomba, QLD 4350 Australia}
\author[0000-0003-3216-0626]{Brett C. Addison}
\affil{University of Southern Queensland, Centre for Astrophysics, West Street, Toowoomba, QLD 4350 Australia}
\author[0000-0002-7830-6822]{Matthew W. Mengel}
\affil{University of Southern Queensland, Centre for Astrophysics, West Street, Toowoomba, QLD 4350 Australia}
\author{Jack Okumura}
\affil{University of Southern Queensland, Centre for Astrophysics, West Street, Toowoomba, QLD 4350 Australia}

\begin{abstract}

We report the discovery of a highly eccentric long-period Jovian planet orbiting the hot-Jupiter host HD\,83443. By combining radial velocity data from four instruments (AAT/UCLES, Keck/HIRES, HARPS, Minerva-Australis) spanning more than two decades, we find evidence for a planet with m~sin~$i=1.35^{+0.07}_{-0.06}$\,\mj, moving on an orbit with $a=8.0\pm$0.8\,au and eccentricity $e=0.76\pm$0.05. We combine our radial velocity analysis with \textit{Gaia} eDR3 /\textit{Hipparcos} proper motion anomalies and derive a dynamical mass of $1.5^{+0.5}_{-0.2} M_{\rm Jup}$. We perform a detailed dynamical simulation that reveals locations of stability within the system that may harbor additional planets, including stable regions within the habitable zone of the host star. HD\,83443 is a rare example of a system hosting a hot Jupiter and an exterior planetary companion. The high eccentricity of HD\,83443c suggests that a scattering event may have sent the hot Jupiter to its close orbit while leaving the outer planet on a wide and eccentric path. 

\end{abstract}



\section{Introduction} \label{sec:intro}

Thirty years ago the only planetary system known was the Solar System.
We now know of thousands of exoplanets\footnote{The NASA Exoplanet Archive, \url{https://exoplanetarchive.ipac.caltech.edu/}, lists 4,569 confirmed exoplanets as of 2021 Nov 15.}, and have learned that such worlds are far more diverse than we ever expected based on our system. Despite this great unveiling, we have yet to discover the true place of the Solar system among its siblings. Is the Solar system's architecture common, or unusual? Are planetary systems like our own rare, or the norm?\footnote{For a detailed overview of our knowledge of the Solar system in the context of exoplanetary science, we direct the interested reader to \citet{SSRev}, and references therein} 

Because of the challenges involved in finding Earth-sized planets on Earth-like orbits around Sun-like stars \citep[e.g.][]{endl15}, a key focus of current research efforts is the search for, and study of, Jupiter- and Saturn analogues -- planets similar to the Solar system's two gas giants, moving on similar, long-period orbits \citep[e.g.][]{zech13,2Jup,rowan16,Analogues2,30177,fulton2021}. The study of such objects can provide vital clues as to the degree to which the Solar system is unusual, helping to place our own planetary system in the context of the wider population. Whilst the scientific value of such Solar system analogues is high, they remain challenging to detect, with the main exoplanet detection methods being strongly biased towards the discovery of planets in systems that are not like our own \citep[see e.g.][and references therein]{exoplanethandbook}.

The formation of planets is a byproduct of the formation of stars. Vast clouds of dust and gas collapse, leading to the formation of a protostar, surrounded by a circumstellar disk from which the planets form \citep[e.g.][]{1997ASPC..122...37B, 2000prpl.conf.....M, 2003ApJ...599..577B}. Initially, those planets
form through the collisional accumulation of successively larger bodies 
\citep{1990AREPS..18..205W}. If this process takes longer than the typical lifetime of the gaseous component of the circumstellar disk, the resulting planets will be comparable to the Solar system's terrestrial planets -- whose accretion likely took around 10$^8$ years to complete \citep[e.g.][]{Ch04,ROB11}. Conversely, if the embryos grow sufficiently massive prior to the dissolution of the gaseous component of the circumstellar disk (reaching $\sim$~10$M_\Earth$), they become capable of feeding on the gas in the disk, allowing them to rapidly 
obtain a gaseous envelope \citep{1974Icar...22..416P, 2007Natur.448.1022J}. This process of rapid, runaway growth continues until the planet has opened a gap in the circumstellar disk, after which the growth slows markedly as the planet feeds from the edges of that gap. Such feeding will often be accompanied by the migration of the resulting giant planet -- a process which can also be driven by the planet's interaction with the other embryos and planetesimals surrounding the star. 

In the Solar system, such migration is suggested by the sculpting of the Asteroid and Edgeworth-Kuiper belts \citep[e.g.][]{Gomes97,MM09,MM11}, and the captured Jovian and Neptunian Trojan populations \citep[e.g.][]{Cap1,LH1,Pir19} -- revealing that giant planets can migrate both inward and outward during the latter stages of their formation. The hundreds of giant exoplanets that have been orbiting closer than Jupiter orbits the Sun, have forced a re-evaluation of planet formation theories. This phenomenon can be explained, at least partially, by the observational bias that makes these short-period giant planets much easier to detect. It is considered likely that these giant, close-in planets have migrated from their initial formation location inwards to the vicinity of their host stars during their accretion 
\citep[e.g.][]{2010AsBio..10...19A, 1986ApJ...309..846L, 1996Natur.380..606L, 1997Icar..126..261W, 1997ApJ...482L.211W, 2002ApJ...565.1257T, 2005A&A...434..343A, 2005ApJ...626L..57A, mord2009a, mord2009b}, though the formation of close-in gas giant planets in-situ via the core-accretion process could still explain the origin of some of these planets \citep[see e.g.,][]{2016ApJ...829..114B,2019A&A...629L...1H}.



Three distinct mechanisms have been proposed to explain the inward migration of giant planets from initially distant orbits to become such ``hot Jupiters'' \citep[e.g. the review by][]{2018ARA&A..56..175D}. The first is a migration process driven by interactions between a young giant planet and the circumstellar disk, as the planet feeds from that disk \citep{1980ApJ...241..425G, 1986ApJ...309..846L, 2014prpl.conf..667B, heller19}. Torques between the planet and the edges of the gap it has opened in the disk can cause the planet to migrate inwards -- a process that either ends when the planet reaches the inner edge of the disk, or when the disk itself is stripped by the youthful star \citep{1996Natur.380..606L}. The migration of the planet in this mechanism is smooth, with the orbital eccentricity and inclination of the planet's orbit (with respect to the disk) remaining essentially zero throughout - leading to hot and warm Jupiters moving on circular orbits that are well aligned with the equatorial plane of their host star \citep{bitsch13}.

The second proposed mechanism invokes dynamical excitation and mutual scattering events between giant planets \citep[e.g.][]{1996Sci...274..954R, 1996Natur.384..619W, scat3, ford08}. In such a scenario, two giant planets experience a close encounter (or a series of such encounters) that act to significantly increase the orbital eccentricities of both planets. One planet is flung inwards, eventually reaching an orbit with a very small periapse. The other planet is flung outwards, onto a highly eccentric orbit with periapsis close to the apoapse of the inner planet's orbit. Tidal forces between the star and the inner planet then act to circularise that planet's orbit at periapsis, decoupling the two planets, and leaving a hot Jupiter on a near circular orbit and a (much) more distant planet with an eccentric orbit \citep[e.g.,][]{nagasawa08}. In extreme cases, this process could even lead to the ejection of the outer planet from the system entirely. The hot Jupiter so produced would be expected to have low or moderate orbital inclination (relative to the equatorial plane of its star), due to the near-coplanarity expected of the initial orbits of the interacting planets.

The final mechanism involves the dynamical interaction of a massive planet and a distant companion on an orbit inclined by $\gtrsim$30$\deg$ with respect to the plane of the planet’s orbit \citep{kozai, lidov}. Over timescales much longer than the inner companion's orbital period, the planet's orbit is perturbed by the distant companion through a series of Kozai-Lidov oscillations, resulting in the coupled evolution of orbital inclination and eccentricity. Such evolution can drive the planet onto a highly eccentric orbit -- at which point tidal interactions between the planet and star at periapsis once again act to circularise its orbit, freezing in the enhanced orbital inclination present at that phase in its cyclical evolution. Hot Jupiters produced in this manner would be expected to display strong orbital misalignment - moving on orbits that are highly tilted, or even retrograde, with respect to the equatorial planes of their host stars \citep[e.g.][]{fw09, winn09, naoz11, dalal19}.

In this context, the study of exoplanetary systems containing hot Jupiters is particularly interesting. It is likely that each of the proposed mechanisms will contribute to the overall population of observed hot Jupiters -- and it is interesting to attempt to disentangle which such planets reached their current orbits as a result of which mechanism.

In this work, we examine the HD\,83443 planetary system which has long been known to host a hot Jupiter with an orbital period of 2.9855$\pm$0.0004\,d and an eccentricity of 0.05$\pm$0.05 \citep{Butler2002}. A recent re-analysis of the complete California Planet Search catalog by \citet{rosenthal21} presented refined parameters of $m$ sin\,$i = 0.409\pm$0.019\,\mj\ and $e=0.074^{+0.031}_{-0.032}$.  HD\,83443 was reported to host two giant planets in the early days of extrasolar planet search \citep{fakeplanet}, with CORALIE data suggesting a candidate second Saturn-mass planet at an orbital period of 29.8 days. Further independent analysis of Keck/HIRES and AAT/UCLES data by \citet{Butler2002} confirmed the hot Jupiter but found no evidence for a second planet. 

Given the lengthy archive of radial velocity data now available for HD\,83443, the system stands as an interesting test case for the various theories of planetary migration. If that migration were the result of tidal interaction with the circumstellar disk, it might be the case that additional giant planets orbit at greater distances, awaiting discovery once sufficient data are available. Equally, if the planet were scattered inwards through encounters with another giant planet, then that planet might remain in the system, moving on a highly eccentric orbit -- a smoking gun for the origin of the hot Jupiter HD\,83443\,b.

In this paper, we report the discovery of HD\,83443c, a rare highly eccentric long-period giant planet orbiting a star with a known hot Jupiter. Section 2 details the observational data, and in Section 3, we describe the properties of the host star. Section 4 gives the results of the radial-velocity fitting, a dynamical investigation of the HD\,83443 system is given in Section 5, and Sections 6 and 7 present our discussion and conclusions.

\section{Observations and Data Reduction}\label{sec:observations}

HD\,83443 has been observed by four precise radial velocity instruments spanning a baseline of over 22 years. Here we give details about the observations from each instrument. All radial velocities used in this analysis are given in Table~\ref{tab:allRV}.


\begin{deluxetable}{cccc}
\tabletypesize{\scriptsize}
\tablecaption{{Radial Velocities for HD\,83443}
\label{tab:allRV}}
\tablehead{
\colhead{Time} & \colhead{Velocity} & \colhead{Uncertainty} & \colhead{Instrument}\\
\colhead{[BJD]} & \colhead{[\mos]} & \colhead{[\mos]} & \colhead{}}
\startdata
\multicolumn4c{{\textsc{Minerva}}-Australis - ThAr} \\
2458523.05882338 & 28316.1248 & 8.88 & {\textsc{Minerva}}-ThAr \\
2458523.06634676 & 28292.2683 & 7.54 & {\textsc{Minerva}}-ThAr \\
2458527.22736930 & 28176.2535 & 10.04 & {\textsc{Minerva}}-ThAr \\
2458527.23836497 & 28196.2810 & 7.24 & {\textsc{Minerva}}-ThAr \\
2458530.13194543 & 28196.9606 & 7.77 & {\textsc{Minerva}}-ThAr \\
2458530.14292949 & 28216.8482 & 9.29 & {\textsc{Minerva}}-ThAr \\
2458532.19110159 & 28248.3593 & 7.11 & {\textsc{Minerva}}-ThAr \\
2458532.20209720 & 28263.8133 & 10.75 & {\textsc{Minerva}}-ThAr \\
2458533.20590349 & 28219.9088 & 13.91 & {\textsc{Minerva}}-ThAr \\
2458534.13895505 & 28283.3215 & 7.60 & {\textsc{Minerva}}-ThAr \\
\enddata
\tablecomments{Table~\ref{tab:allRV} is published in its entirety in machine-readable format online. A portion is shown here for guidance regarding its form and content.}
\end{deluxetable}

\subsection{AAT}
The Anglo-Australian Planet Search (AAPS) survey started in 1998. The survey was carried out using the 3.9~m Anglo-Australian Telescope (AAT) and the University College London Echelle Spectrograph (UCLES) with a limiting Doppler precision of 3 m$s^{-1}$ \citep{2001ApJ...551..507T}. AAPS obtained 25 observations of HD\,83443 between UT 1999 Feb 2 and UT 2015 Mar 13.

\subsection{HARPS}
HARPS is a highly-stabilised spectrometer which began operations in 2003 \citep{2003Msngr.114...20M}.  The HARPS fibre feed was upgraded in 2015 \citep{fibre}; here we use 45 observations taken prior to this correction, from UT 2003 Dec 28 to 2015 May 1, and 11 observations that were obtained afterward, from 2015 Dec 10 to 2016 May 28.  We use the HARPS radial velocities from  \citet{trifonov2020public}, which corrected for nightly zero point and CCD stitching offsets. 

\subsection{Keck/HIRES}
The Keck/HIRES Radial Velocity Survey started in 1994, and focused on the search for exoplanets around low activity F, G, K and M-dwarf stars. This program acquired more than 60,000 radial velocity measurements of 1,624 stars \citep{2017AJ....153..208B}. A revised data release, correcting for small zero-point offsets and other systematics, was published in
\citet{tal2019correcting}. From this database we obtain 45 observations of HD\,83443, taken between UT 2000 Dec 19 and 2014 Dec 11.

\subsection{{\textsc{Minerva}}-Australis}

\textsc{Minerva}-Australis saw first light in 2019, and is a dedicated facility that spends every clear night obtaining radial velocity measurements of stars thought to be potential planet hosts \citep{2018arXiv180609282W,addison2019,TOI257}.
\textsc{Minerva}-Australis consists of an array of four independently operated 0.7\,m Planewave CDK700 telescopes situated at the Mount Kent Observatory in Queensland, Australia \citep{addison2019}. Each telescope simultaneously feeds stellar light via fibre optic cables to a single KiwiSpec R4-100 high-resolution ($R=80,000$) spectrograph \citep{2012SPIE.8446E..88B} with wavelength coverage from 480 to 620\,nm. 

A total of 22 individual spectra for HD\,83443 were obtained between 2019 Feb 8 and 2021 Feb 22 using {\textsc{Minerva}}-Australis telescope 4. Of these, 17 epochs used the simultaneous Th-Ar calibration fibre, and 5 epochs used a simultaneous back-illuminated iodine cell for wavelength calibration. The latter technique was employed from 2019 Dec 4 to mitigate the loss of spectral information due to saturation from the argon lines. Radial velocities were derived by cross-correlation, where the template being matched is the mean spectrum.

\section{Stellar Properties of HD83443}\label{sec:thestar}

Table~\ref{tab:star} summarises the literature measurements for the properties of HD\,83443. HD\,83443 is a solar mass K0 star, with more than twice the metallicity of the Sun. This enhanced metallicity is consistent the well-established giant planet-metallicity correlation \citep[e.g.][]{fv05, jones16, witt17, ghezzi18, osborn2020, fulton2021}. For the orbit fitting and derivation of planetary parameters, we adopt a stellar mass of 1.00$\pm$0.03\,\msun \citep{Delgado_Mena_2019}.

\begin{deluxetable}{lccc}
\tablewidth{0.45\textwidth}
\tablecolumns{4}
\tablecaption{Stellar parameters for HD\,83443. The adopted stellar mass is indicated in bold. \label{tab:star}}
\tablehead{
\colhead{Parameter} & \colhead{Value} & \colhead{Reference} \\
}
\startdata
Right Ascension (h:m:s)  & 9:37:11.8276 & 1 \\ \hline
Declination (d:m:s)    & -43:16:19.9326 & 1 \\ \hline
Distance (pc)       & 40.95~$\pm$~0.06 & 2 \\ \hline
Spectral type       & K0 \textrm{V} & 3 \\ \hline
$(B - V)$         & 0.811 & 4 \\ \hline
$T_{\rm eff}$ (K)     & 5429$^{+96}_{-122}$ & 5 \\
             & 5487$^{+90}_{-107}$ & 6 \\
             & 5442~$\pm$~17 & 7 \\ \hline
$\log g$ (cm$^{2}$/s)   & 4.41$^{+0.08}_{-0.07}$ & 5 \\ 
             & 4.39~$\pm$~0.04 & 7 \\
             & 4.43~$\pm$~0.08 & 8 \\ \hline
$R_{\star}$ ($R_{\odot}$) & 1.005~$^{+0.055}_{-0.038}$ & 5 \\
             & 0.982$^{+0.039}_{-0.031}$ & 6 \\
             & 0.94~$\pm$~0.02 & 7 \\ \hline
$L_{\star}$ ($L_{\odot}$) & 0.790$^{+0.020}_{-0.015}$ & 5 \\ 
             & 0.787~$\pm$~0.002 & 6 \\ 
             & 0.72$^{+0.12}_{-0.10}$ & 9 \\ \hline
$M_{\star}$ ($M_{\odot}$) & 0.95~$\pm$~0.12 & 5 \\
             & 0.79~$\pm$~0.07 & 7 \\
             & 1.05~$\pm$~0.10 & 9 \\ 
             & 1.00~$\pm$~0.03 & 12 \\ \hline
Metallicity, [Fe/H]    & 0.34~$\pm$~0.03 & 8 \\
             & 0.44~$\pm$~0.04 & 9 \\
             & 0.35~$\pm$~0.08 & 11 \\ \hline
Age (Gyr)         & 3.2 & 4 \\ 
             & 2.64$~\pm$~2.49 & 12 \\ \hline
$\rho_{\star}$ (g cm$^-3$) & 1.32$\pm$0.28 & 5 \\ \hline
$v \sin i$ (\kms)     & 1.4    & 4 \\ 
             & 1.3~$\pm$~0.5 & 10 \\ \hline
$P_{rot}$ (day)      & 35.3 & 4 \\ \hline
\enddata
\raggedright
\tablerefs{1. \cite{gaia2018}; 2. \cite{2018yCat.1345....0G}; 3. \cite{1978mcts.book.....H}, 4. \cite{mayor04}, 5. \cite{2019AJ....158..138S}, 6. \cite{gaia2018}, 7. \cite{2017AJ....153..136S}, 8. \cite{sousa08}, 9. \cite{ 2010ApJ...720.1290G}, 10. \cite{2005ApJS..159..141V}, 11. \cite{santos04}, 12. \citet{Delgado_Mena_2019} }

\end{deluxetable}

\section{Orbit Fitting and Results}\label{sec:Results}

While HD\,83443 was part of the main Anglo-Australian Planet Search for 16 years \citep{cooljupiters}, the sparse sampling prevented the detection of longer-period radial velocity signatures. Inspection of additional publicly available data (CORALIE, HARPS, HIRES) with the Geneva group's DACE tool\footnote{ \url{https://dace.unige.ch/dashboard/}} revealed evidence for a potential high-eccentricity, long-period signal. We then performed some initial simple analysis with \textit{Systemic Console 2.200} on the Keck/HIRES, HARPS and AAT data sets (see Section \ref{sec:observations}). Those initial efforts supported the existence of a highly eccentric planet with a period of $\sim$6800 or $\sim$10000 days. 

\begin{figure} 
 \begin{center}
 \begin{tabular}{cc}

  \includegraphics[scale=0.41]{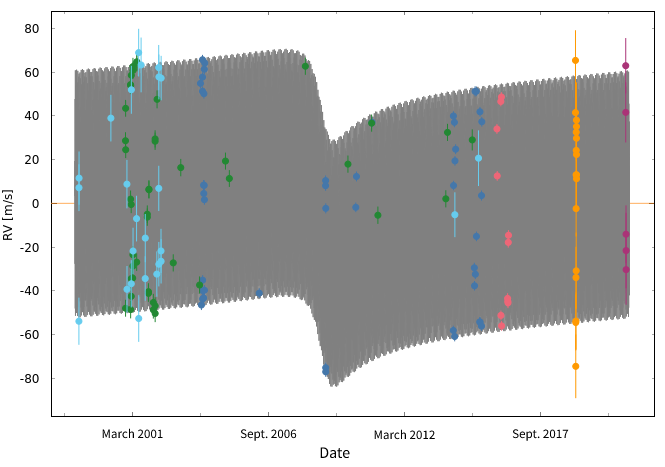} & 
  \includegraphics[scale=0.4]{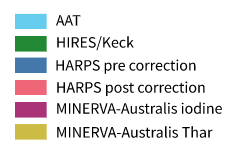} \\
  \includegraphics[scale=0.54]{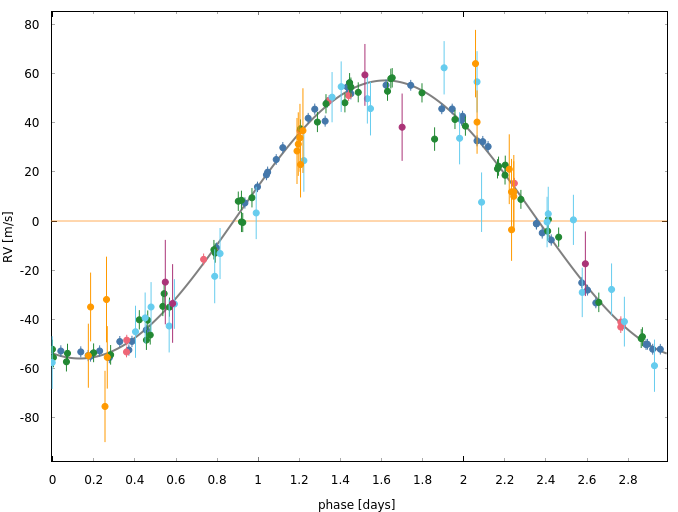} & \\ 
  \includegraphics[scale=0.273]{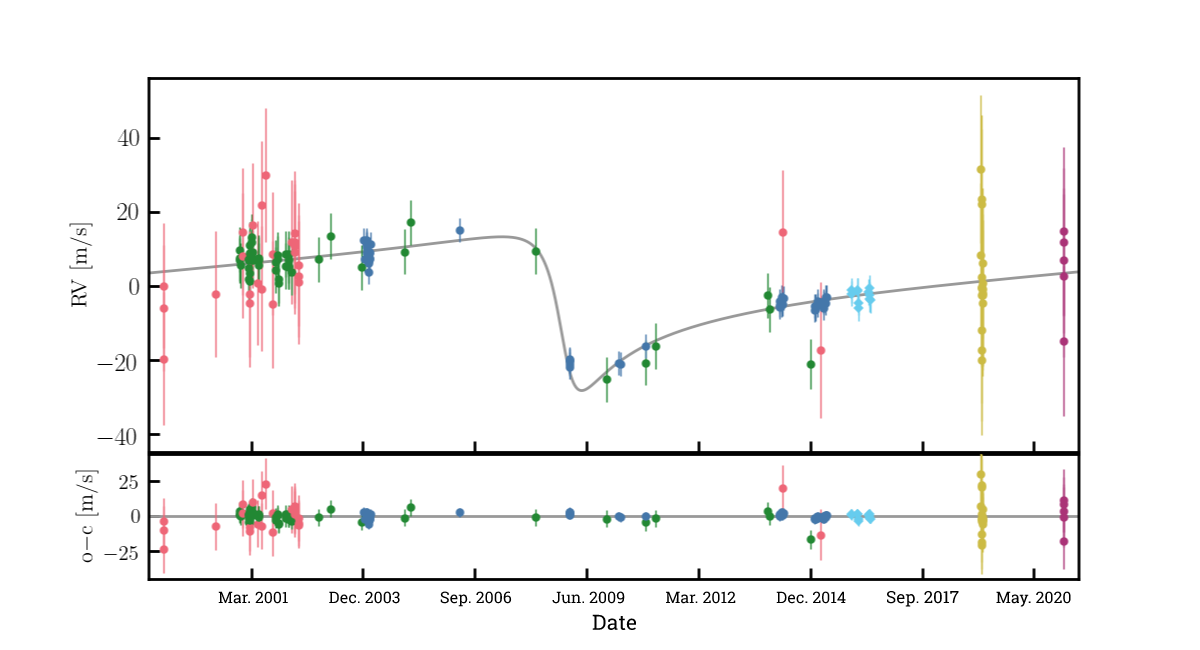} & \\
  
  \end{tabular}
  \caption{The observational data for HD\,83443, colour coded as follows: Light blue - AAT \citep{2001ApJ...551..507T}, Green - HIRES/Keck \citep{tal2019correcting}, Blue - HARPS pre-correction \citep{trifonov2020public}, Red - HARPS post correction \citep{trifonov2020public}, Orange - {\textsc{Minerva}}-Australis ThAr \citep{addison2019} and Purple - {\textsc{Minerva}}-Australis iodine \citep{addison2019}. The top panel shows all the data as a function of time, with the two planet solution shown in black. The centre panel shows the data phase folded for HD\,83443~b only (centre panel), with the lower panel showing the data for the long period planet, HD\,83443~c (lower panel).} \label{fig:RV_phase_folding}
 \end{center}
\end{figure}

For the final fitting, we performed two runs with \textit{Exostriker} \citep{2019ascl.soft06004T}, with initial values for the period of the outer planet at 6000 days and 10000 days. All other priors and starting values were identical between the two runs (Table~\ref{tab:MCMC_param}). The radial velocity time series, the phase-folded plot for the inner planet and the linear time plot for the outer one can be seen in figure \ref{fig:RV_phase_folding}.

We explore the parameter space running a Markov Chain Monte Carlo sampling of the posterior solutions using \texttt{emcee} \citep{Foreman-Mackey_2013} with 88 walkers and 5000 steps. Corner plots for both runs can be seen in the Appendix \ref{appendix}; the posteriors are unimodal and consistent with a highly significant and unique solution.  Posteriors and results are included in Table \ref{tab:MCMC_param}.  Some statistical values related to both runs with \textit{Exostriker} are included in table \ref{tab:stats}. Both \textit{Exostriker} runs converged on essentially identical solutions. The results support the presence of a long-period, highly eccentric outer planet.  The critical periastron velocity excursion is sampled independently by HARPS and Keck/HIRES, giving further support to the interpretation as Keplerian orbital motion.

\begin{table*}[ht]

\centering  
\caption{{MCMC sampling posteriors and priors of the orbital and 
nuisance parameters of the HD\,83443 system, derived by radial velocities (MINERVA, AAT, HARPS, HIRES/Keck). Derived jitter values are simply added in quadrature to the reported error bars.
}}
\label{tab:MCMC_param}

\begin{tabular}{p{6.95cm} r r r r p{6.95cm} p{7.95cm} p{3.95cm} p{3.95cm} p{3.95cm} p{3.95cm} rrrrrrr }   

\hline\hline \noalign{\vskip 0.7mm}

 Parameter & Median and $1\sigma$ & Adopted priors \\ \noalign{\vskip 0.9mm}
\hline
 
$K_b$ [m\,s$^{-1}$]      &56.73$_{-0.36}^{+0.36}$ & $\mathcal{U}(0,10^4)$  \\ \noalign{\vskip 0.9mm}

$P_b$ [day] & 2.985628$_{-0.000005}^{+0.000005}$ & $\mathcal{U}(0,10^5)$  \\ \noalign{\vskip 0.9mm}

$e_b$              &0.012$_{-0.006}^{+0.007}$ & $\mathcal{U}$(0,1)  \\ \noalign{\vskip 0.9mm}

$\omega_b$ [deg]        & 344$_{-15}^{+11}$ & $\mathcal{U}$(0,360)  \\ \noalign{\vskip 0.9mm}
               
$M_{\rm 0,b}$ [deg]      & 335$_{-11}^{+15}$ & $\mathcal{U}$(0,360) \\ \noalign{\vskip 0.9mm}

 $a_b$ [au]           & 0.0406$\pm$0.0004 & (derived)  \\ \noalign{\vskip 0.9mm}

m $\sin i_b$ [$M_{\rm jup}$]     & 0.402$\pm$0.008 & (derived)  \\ \noalign{\vskip 0.9mm}
\hline

$K_c$ [m\,s$^{-1}$]      & 20.7$_{-1.6}^{+2.2}$ & $\mathcal{U}(0,10^4)$  \\ \noalign{\vskip 0.9mm}

$P_c$ [day]          & 8241$_{-530}^{+1019}$ & $\mathcal{U}(0,10^5)$  \\ \noalign{\vskip 0.9mm}

$e_c$              & 0.760$_{-0.047}^{+0.046}$ & $\mathcal{U}$(0,1)  \\ \noalign{\vskip 0.9mm}

$\omega_c$ [deg]        & 118.2$_{-5.7}^{+5.6}$ & $\mathcal{U}$(0,360)  \\ \noalign{\vskip 0.9mm}
               
$M_{\rm 0,c}$ [deg]      & 281.0$_{-4.9}^{+7.9}$ & $\mathcal{U}$(0,360) \\ \noalign{\vskip 0.9mm}

 $a_c$ [au]           & 8.0$\pm$0.8 & (derived)  \\ \noalign{\vskip 0.9mm}

m $\sin i_c$ [$M_{\rm jup}$]     & 1.35$^{+0.07}_{-0.06}$ & (derived)  \\ \noalign{\vskip 0.9mm}
               
\hline

RV off.$_{\rm HARPS pre-correction}$ [m\,s$^{-1}$] & -0.8$_{-0.5}^{+0.5}$ & $\mathcal{U}(-10^5,10^5)$  \\ \noalign{\vskip 0.9mm} 
                   
RV off.$_{\rm HARPS post-correction}$ [m\,s$^{-1}$] & 4.9$_{-1.0}^{+0.9}$ & $\mathcal{U}(-10^5,10^5)$  \\ \noalign{\vskip 0.9mm} 
                   
RV off.$_{\rm HIRES/Keck}$ [m\,s$^{-1}$] & -5.7$_{-0.6}^{+0.6}$ & $\mathcal{U}(-10^5,10^5)$  \\ \noalign{\vskip 0.9mm} 

RV off.$_{\rm AAT}$ [m\,s$^{-1}$] & 9.0$_{-2.2}^{+2.2}$ & $\mathcal{U}(-10^5,10^5)$  \\ \noalign{\vskip 0.9mm} 

RV off.$_{\rm \textsc{Minerva}-Australis iodine}$ [m\,s$^{-1}$] & 28205.9$_{-7.2}^{+7.2}$ & $\mathcal{U}(-10^5,10^5)$  \\ \noalign{\vskip 0.9mm}  
                   
RV off.$_{\rm \textsc{Minerva}-Australis ThAr}$ [m\,s$^{-1}$] & 28250.8$_{-3.4}^{+3.5}$ & $\mathcal{U}(-10^5,10^5)$  \\ \noalign{\vskip 0.9mm}                    
RV jitter$_{\rm HARPS pre-correction}$ [m\,s$^{-1}$] & 1.9$_{-0.3}^{+0.3}$ & $\mathcal{U}(0,10^4)$  \\ \noalign{\vskip 0.9mm} 
                   
RV jitter$_{\rm HARPS post-correction}$ [m\,s$^{-1}$] & 1.9$_{-0.5}^{+0.7}$ & $\mathcal{U}(0,10^4)$  \\ \noalign{\vskip 0.9mm} 
                   
RV jitter$_{\rm HIRES/Keck}$ [m\,s$^{-1}$] & 3.5$_{-0.5}^{+0.6}$ & $\mathcal{U}(0,10^4)$  \\ \noalign{\vskip 0.9mm} 
                   
RV jitter$_{\rm AAT}$ [m\,s$^{-1}$] & 9.8$_{-1.7}^{+2.2}$ & $\mathcal{U}(0,10^4)$  \\ \noalign{\vskip 0.9mm}

RV jitter$_{\rm \textsc{Minerva}-Australis iodine}$ [m\,s$^{-1}$] & 10.2$_{-6.3}^{+10.9}$ & $\mathcal{U}(0,10^4)$  \\ \noalign{\vskip 0.9mm} 
                   
RV jitter$_{\rm \textsc{Minerva}-Australis ThAr}$ [m\,s$^{-1}$] & 10.3$_{-3.5}^{+4.1}$ & $\mathcal{U}(0,10^4)$  \\ \noalign{\vskip 0.9mm}                    
\hline \noalign{\vskip 0.7mm}
\tablecomments{$\mathcal{U}$(l,u) signifies a uniform prior with lower bound $l$ and upper bound $u$.}
\tablecomments{Mean anomalies $M_{\rm 0}$ are for epoch BJD 2453001.85.}
\end{tabular}
\end{table*}

\begin{deluxetable}{ccccccccc}
\tablewidth{0.45\textwidth}
\tablecolumns{9}
\tablecaption{Statistical values for \textit{Exostriker} runs. \label{tab:stats}}
\tablehead{
\colhead{Period (days)} & \colhead{Ecc} & \colhead{Chi} & \colhead{RMS} & \colhead{WRMS} & \colhead{lnL} & \colhead{BIC} & \colhead{AIC} & \colhead{Initial period (days)} \\
}
\startdata
8240 & 0.7598 & 1.0442 & 7.01 & 3.91 & -418.07 & 946.08 & 792.14 & 6000 \\ \hline
8203 & 0.7590 & 1.0923 & 7.03 & 3.93 & -419.72 & 949.38 & 795.45 & 10000 \\ \hline
\enddata
\raggedright
\end{deluxetable}

\section{Dynamical Simulations and Limits on Additional Planets} \label{sec:Dynamics}

The two planets described orbiting HD\,83443 in this work are sufficiently widely separated to be effectively dynamically decoupled from one another. As such, it seems likely that there is a vast amount of space in the system that could host additional planets -- so long as such planets were not disrupted or ejected by whatever process led to the highly eccentric orbit of HD\,83443 c, and the inward migration of HD\,83443 b. 

\subsection{The Dynamical Stability of the HD\,83443 Planetary System}

In order to explore this possibility, we performed a suite of $n$-body simulations, using the Hybrid integrator within the \textsc{Mercury} integration package \citet{mercury}. 
We use a similar methodology to that laid out in \citet{PiMensa}, distributing a large population of massless test particles between the orbits of the two planets, and following their dynamical evolution for a period of one million years. Due to the small orbital distance of HD\,83443 b and the test particles used in the inner regions of the system, a purely Newtonian approach to the dynamical simulations would be inappropriate. We therefore used a version of \textsc{Mercury} modified to include the first-order post-Newtonian relativistic corrections, as described in \citet{Milankovic}, following \citet{Gilmore08}. Test particles were removed from the simulations upon colliding with either of the planets in the system, the central object, or upon reaching a distance of 20 au from the system barycentre (and hence having been transferred to a chaotic orbit crossing that of HD\,83443 c). The simulation timestep was 0.08 days, and the entire suite of simulations took approximately six months to run, distributed across the nodes of USQ's \textit{Fawkes} supercomputing cluster - equivalent to approximately 250 years of CPU time.

Test particles were distributed in a regular grid in $\textrm{log} a - e - \omega -M$ space, with the initial orbital inclinations of both the planets and the test particles set to zero\footnote{In other words, the system was modelled under an implicit assumption of initial coplanarity, on the grounds that no information was available on the mutual inclination of the two known planets.}. In total, 13.6 million test particles were created. 492 unique values of semi-major axis were used as initial conditions for those test particles, evenly distributed in log $a$ space between 0.04095765 and 8.04146231 au. At each of those 492 unique semi-major axes, 123 unique orbital eccentricities were used, evenly distributed in the range 0.0 to 0.9. At each of these 60516 $a-e$ pairs, 45 unique values of $\omega$ were chosen, evenly distributed between 0 and 360$^\circ$. Finally, at each of these 2723220 $a-e-\omega$ locations, 5 unique values of mean anomaly were tested - again evenly distributed between 0 and 360$^\circ$. The orbits and masses of the two planets were taken directly from Table~\ref{tab:MCMC_param}, and are therefore the minimum feasible masses for the planets based on the radial velocity observations. Since the simulations assume the system to initially be coplanar, this seems a reasonable assumption to make. 

\begin{figure} 
 \begin{center}
 \begin{tabular}{cc}
  \includegraphics[scale=0.435]{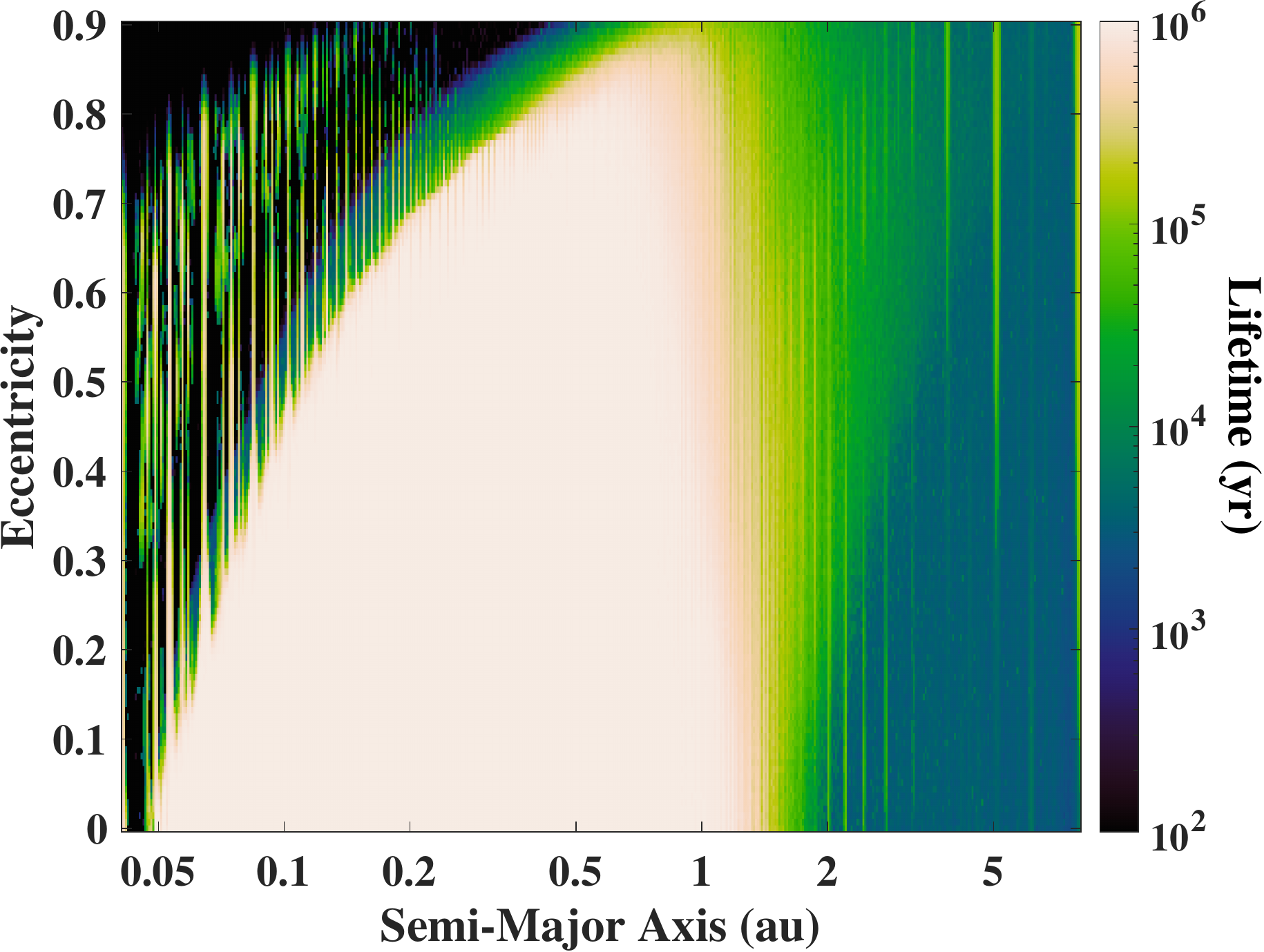} & \includegraphics[scale=0.144]{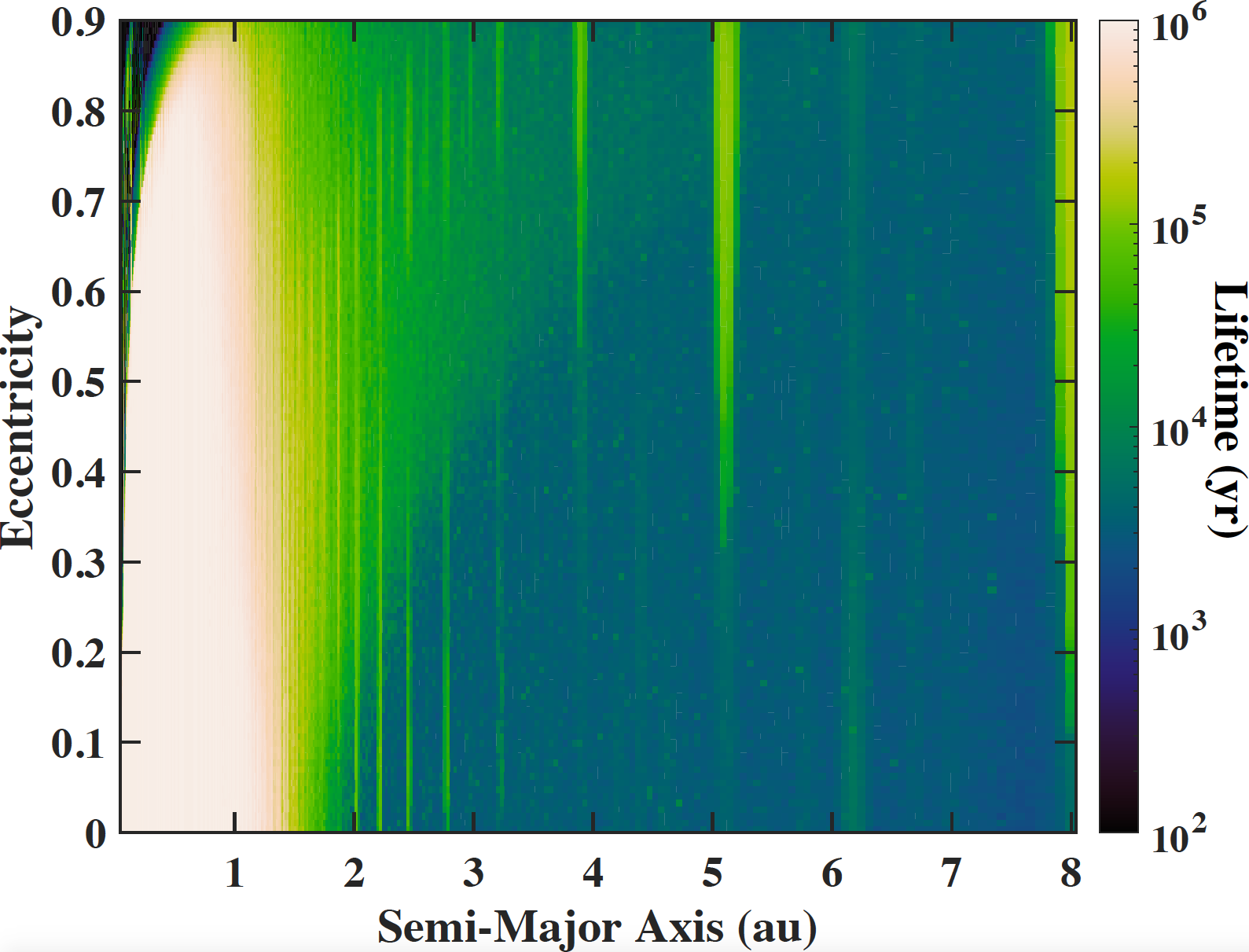} \\
  \end{tabular}
  \caption{The dynamical stability of massless test particles in the HD\,83443 system, as a function of their initial semi-major axis and eccentricity, from simulations spanning a period of 1 Myr. The stability of the inner region of the system is dominated by the influence of HD\,83443~b, with test particles that approach that planet too closely being removed from the system on short timescales. In the outer regions of the system, stability is strongly influenced by HD\,83443~c, which renders most orbits beyond $\sim$1.5 au unstable due to its high eccentricity (reaching a periastron distance of 1.92\,au). We note that, since the simulations covered a period of 1 Myr, the regions in white are those where test particles have minimum lifetimes of 1 Myr. It is highly likely that, across much of that space, lifetimes would be far longer than the 1 Myr of our simulations.}   \label{stab}
 \end{center}
\end{figure}

Of the 13.6 million test particles whose orbital evolution was followed in our simulations, just over 7.9 million were removed from the system within the 1 Myr of our integrations, with particles colliding with one or other of the two planets ($\sim$2.8 million), falling into the central star ($\sim$430000) or being ejected beyond a barycentric distance of 20 au ($\sim$4.7 million). The time at which each of those $\sim$ 7.9 million test particles was removed from the system was recorded, and the results used to generate the stability plots shown in Figure~\ref{stab}. In each of the two panels in that figure the lifetime shown at a given $a$-$e$ location is the mean of the ejection times across the 225 test particles that began the simulation at that particular $a$-$e$ location. 

It is immediately apparent from Figure~\ref{stab} that the vast majority of the 5.7 million test particles that survived the full 1 Myr duration of the simulations were located interior to $\sim$ 1.2 au, forming a broad island of stability bounded, at the inner edge, by orbits that approach HD\,83443\,b (i.e. periastra of $\sim$ 0.05 au), and at the outer edge, by orbits with apastra of $\sim$ 1.2 au\footnote{It should be noted, here, that just because a region would be stable for a particle of minimal or zero mass, that will remain true for objects of arbitrarily large mass. Clearly, at any given location, there is a maximum mass for which such an orbit would remain stable. However, our simulations nonetheless serve as a good guide to the regions of the system that are definitely unstable, from the point of view of additional companions.}.  The left hand panel of Figure~\ref{stab} plots semi-major axes logarithmically, which allows fine structures in the inner region to be clearly seen. In that inner 'unstable wedge', carved by HD\,83443\,b, there are a large number of narrow strips of stability rising to large eccentricities, at the locations of mean-motion resonances between test particles and HD\,83443\,b. 

Perhaps the most pronounced of these resonant features falls right at the inner edge of the plot, with a large number of stable test particles trapped in 1:1 mean-motion resonance with the innermost planet. Such particles are analogues of the Solar system's Jovian and Neptunian Trojan populations (see e.g. \citet{Troj1,Troj2,QR322,Holt21}, and \citet{SSRev} for a more general overview) - populations of asteroids trapped in the same mean-motion resonance, moving on stable orbits that librate around the leading and trailing Lagrange points (L$_4$ and L$_5$) in the orbits of the giant planets. In the Solar system, the populations of stable Trojan companions to the giant planets are thought to have been captured during the final stages of planetary migration \citep[e.g.][]{Cap1,LH1,Cap2} - and the presence of those stable Trojan test particles in our runs serves as a reminder that there is the possibility that the inward migration of giant planets could lead to the capture and transport of Trojans in exoplanetary system \citep[as discussed in e.g.][]{ExoTroj1,ExoTroj2,ExoTroj3,ExoTroj4,ExoTroj5}.

The large number of narrow resonant strips of stability found in this region is the direct result of the high density of test particles deployed in these runs - sampling the space in the inner wedge with sufficient precision that we capture as many resonant features as possible - rather than having test particles distributed such that they can skip resonant orbits at the time of creation. Whilst such oversampling might seem excessive, the presence of so many potentially stable resonant scenarios (particularly those where the stable feature extends down to circular orbits) fits well with the discovery of a number of tightly dynamically packed exoplanetary systems in recent years \citep[e.g.][]{res1, res2, res3, res4}.

Since HD 83443\,c is not known to transit the disk of its host star, its orbital inclination remains unconstrained. As such, it is interesting to consider what impact an inclined orbit for HD 83443\,c might have on the stability of the rest of the planetary system. To investigate this, we therefore carried out five ancillary sets of simulations, to consider the impact of the inclination of the orbit of HD 83443\,c relative to that of the inner planet and disk of test particles. Those simulations used identical suites of test particles, again distributed between the orbits of the two planets. The initial orbits of those test particles were co-planar with the orbit of HD 83443\,b. In total, we simulated the evolution of 686,340 test particles in each scenario. The evolution of those test particles was again followed for a period of one million years, or until they collided with one of the system's massive objects, or were ejected. From one simulation to the next, the only changes we made to the initial conditions were to increase the orbital inclination of HD 83443\,c, testing inclinations of 5, 15, 30, 45, and 60 degrees to the orbit of the innermost planet and test particle disk (in a manner similar to that used in e.g. \citet{HUAqr} and \citet{QSVir}). For each orbital inclination, we recalculated the mass of HD 83443\,c, increasing it such that the M$\sin i$ value of that mass was held constant. As the inclination of the planet's orbit increased, therefore, so did the mass of the planet in our simulations.

\begin{figure} 
 \begin{center}
 \begin{tabular}{ccc}
  \includegraphics[scale=0.33]{HD83443_v1.pdf} & 
  \includegraphics[scale=0.33]{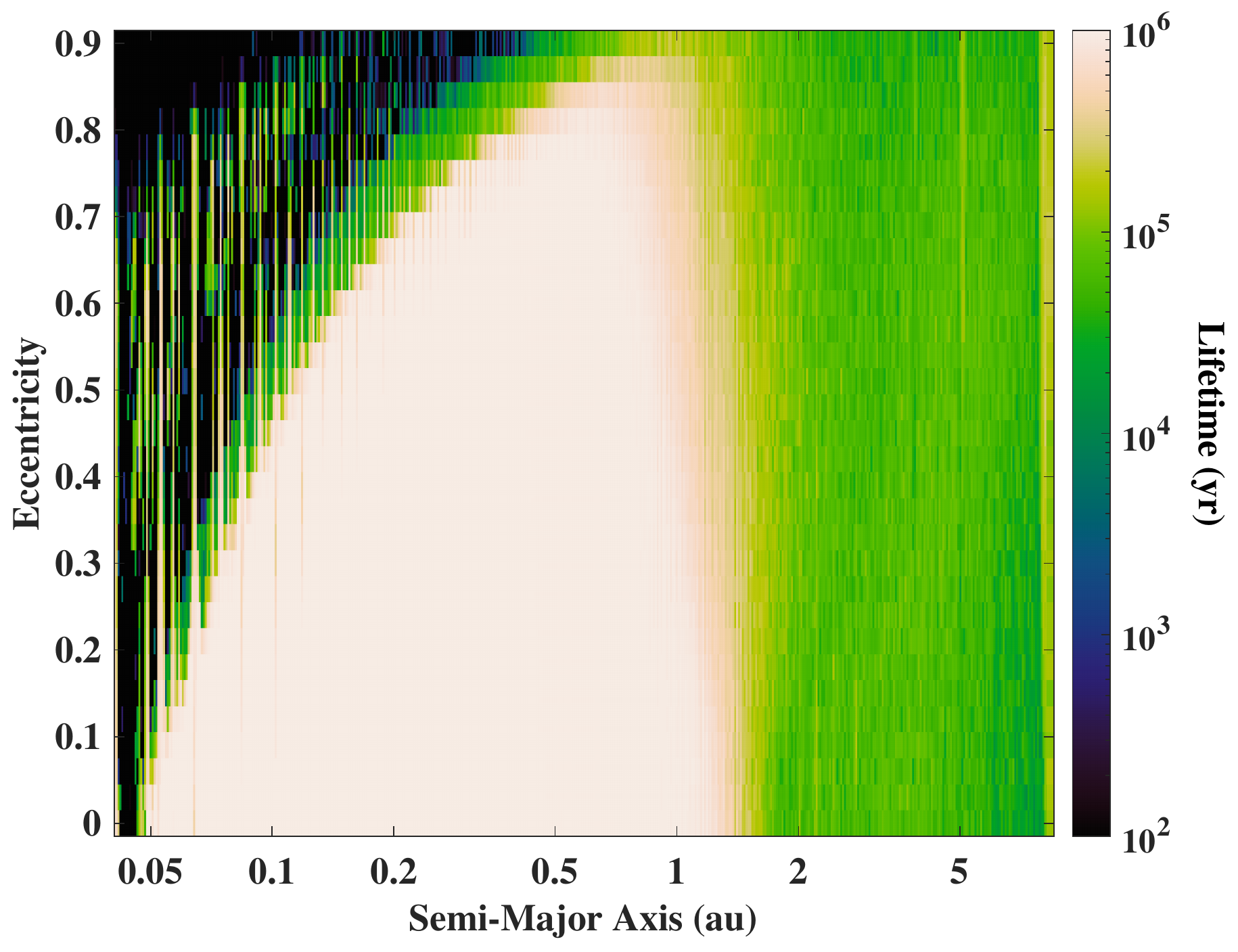} &
  \includegraphics[scale=0.33]{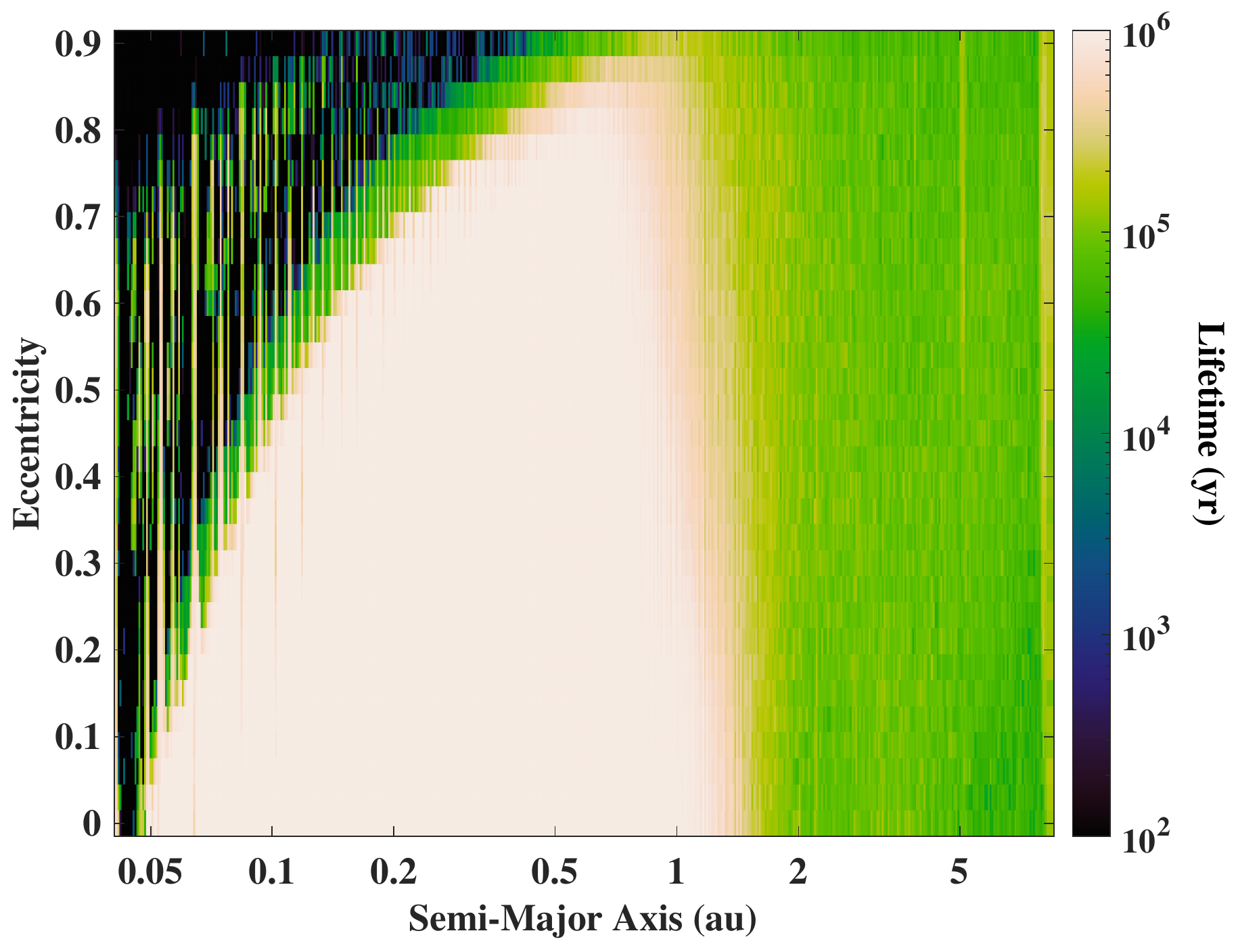} \\
  \includegraphics[scale=0.33]{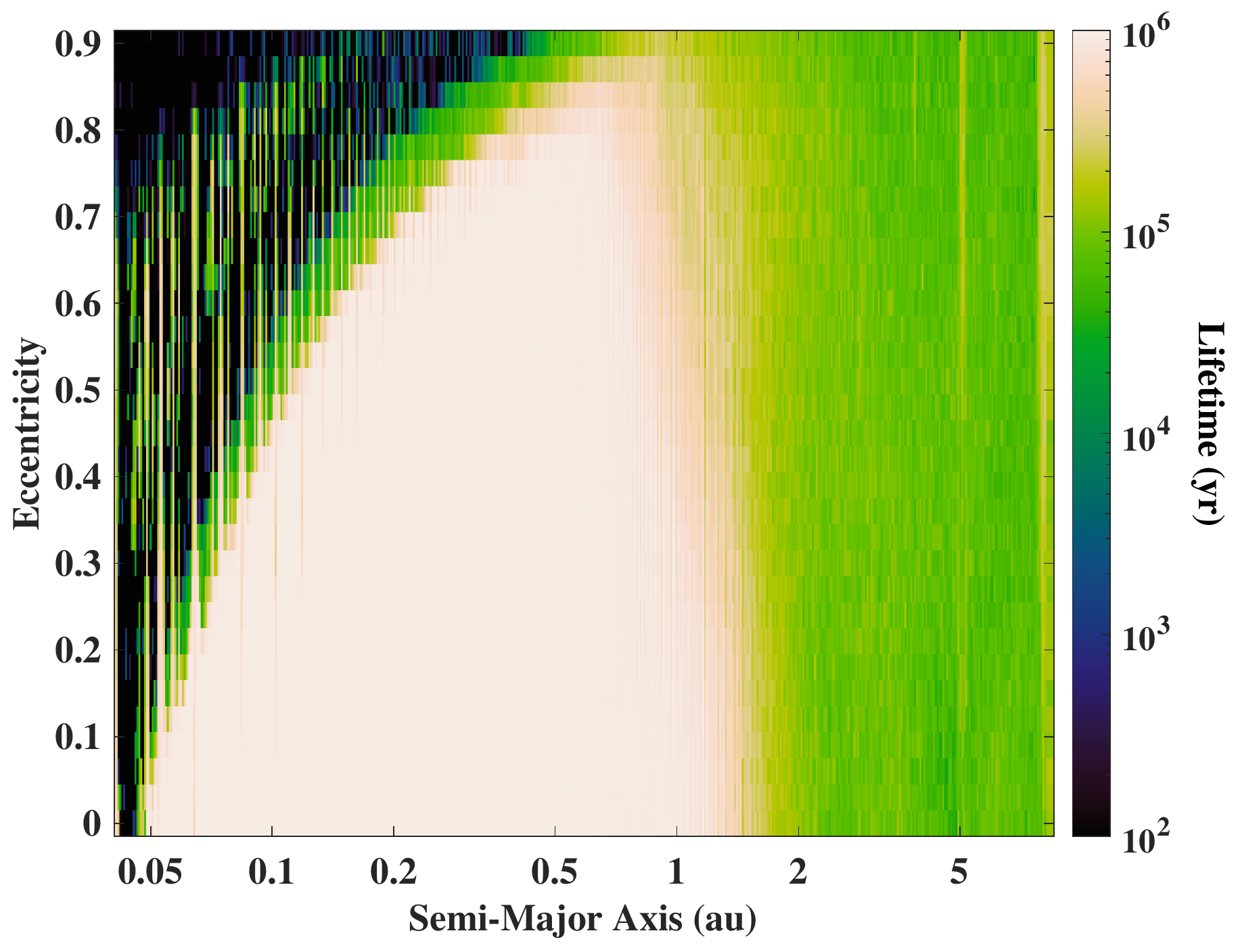} &
  \includegraphics[scale=0.33]{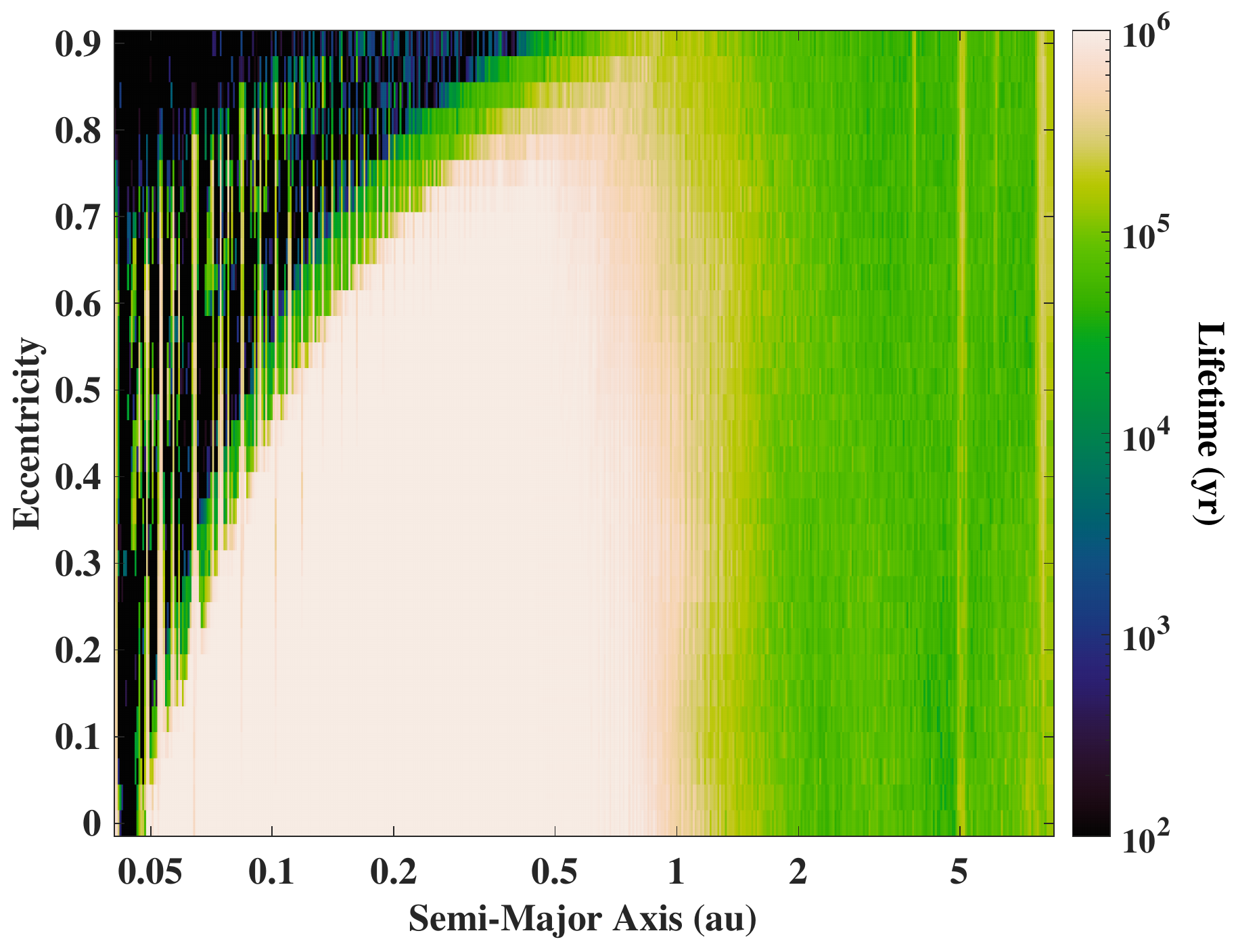} &
  \includegraphics[scale=0.33]{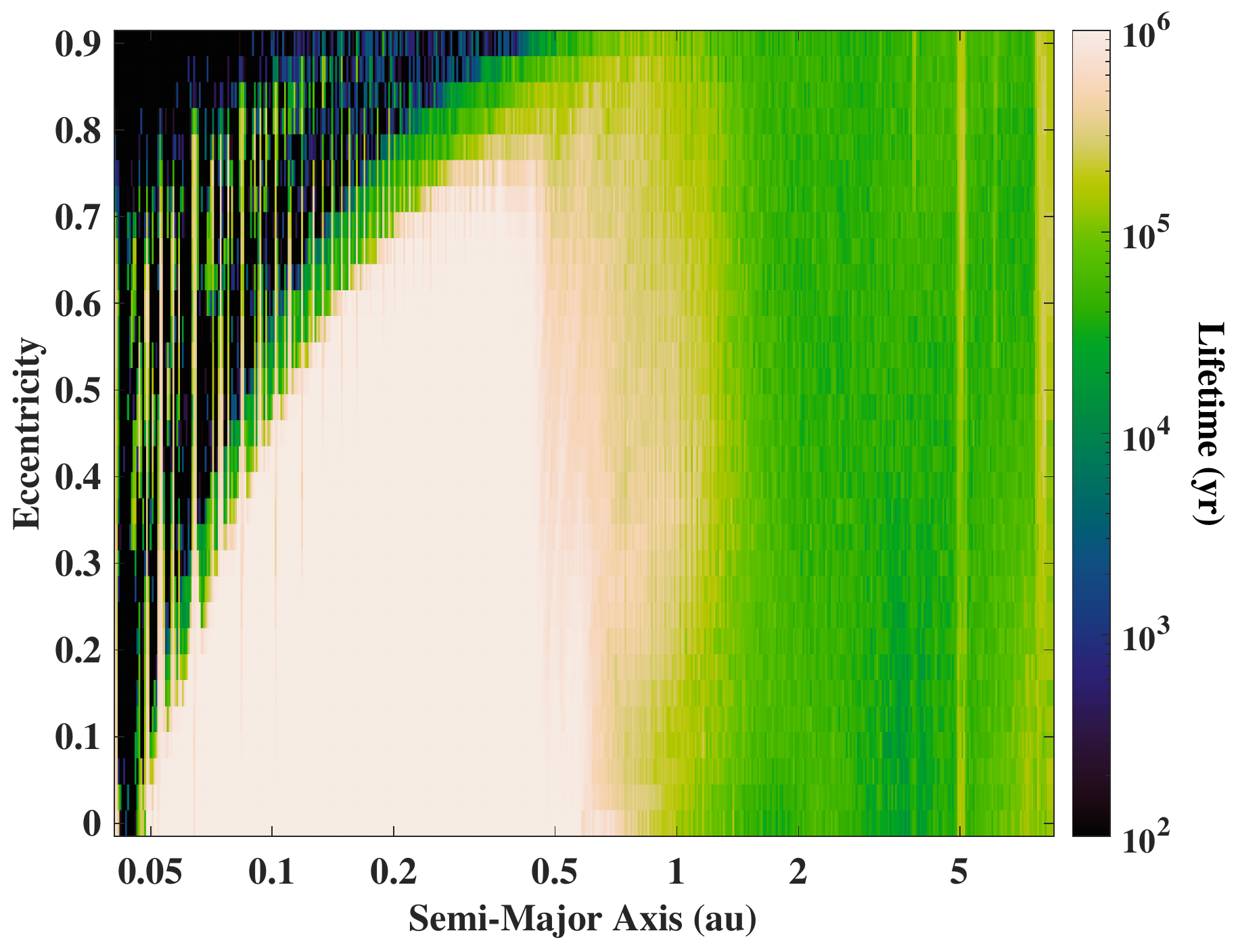} \\
  \end{tabular}
  \caption{The dynamical stability of massless test particles in the HD\,83443 system, for simulations considering a variety of orbital inclinations for the orbit of HD 83443\,c. The top left hand panel shows the case where the system begins in a fully co-planar state, with HD 83443\,c moving on an orbit that is in the same plane as the disk of test particles and the orbit of HD 83443\,b, as presented in Figure~\ref{stab}. The remaining panels feature simulations at lower resolution, following the evolution of 686,340 massless test particles. The central plot of the top row shows the outcome with HD 83443\,c moving on an orbit tilted by 5 degrees to the disk and inner planet; top right has c on an orbit inclined by 15 degrees; lower left has an inclination of 30 degrees; lower centre has an inclination of 45 degrees, and lower right, 60 degrees. As with Figure~\ref{stab}, the simulations spanned a period of 1 Myr. Again, since the simulations covered a period of 1 Myr, the regions in white are those where test particles have minimum lifetimes of 1 Myr. It is highly likely that, across much of that space, lifetimes would be far longer than the 1 Myr of our simulations. Moderate levels of mutual inclination between the orbits of the outer planet and the disk do little to alter the stability profile, but once the orbit of HD 83443\,c is inclined by 45 and 60 degrees, it begins to whittle away at the outer edge of the stable area, rendering orbits in the region beyond $\sim$ 0.9 au (45 degree) and $\sim$ 0.5 au (60 degree) unstable on timescales of hundreds of thousands of years.}   \label{stabinclined}
 \end{center}
\end{figure}

The results of these simulations are shown in Figure~\ref{stabinclined}. As the inclination of HD 83443\,c is increased, the outer boundary of the stable region (which, in the main runs, was located at $\sim$ 1.2 au) slowly moves to smaller semi-major axes. This effect is most pronounced for the two most inclined simulation sets - with the outer planet inclined by 45 and 60 degrees. As the mass of HD 83443\,c is increased between runs, the breadth of the unstable region also increases. For the coplanar scenario, the outer edge of the stable region is located at $\sim$ 1.2 au, which is 5 Hill radii interior to the location of the planet at periastron. At an inclination of 60 degrees, the mass has only doubled. Since the Hill radius of a planet is proportional to the cube root of the planet's mass, this doubling in mass only increases the size of that planet's Hill sphere by approximately 26 \%. At periastron, this corresponds to shifting the outer edge of the stable region to $\sim$ 1 au. In actuality, our simulations show that the unstable region for the 45 degree and 60 degree simulations stretches farther inward than would be expected based solely on the width of HD\,83443\,c's Hill sphere. This is the direct result of the increased inclination of the outer planet, which acts to drive the orbital eccentricity and inclination of test particles in the outer region of the disk to vary, driving the apastra of those test particles to locations within the dynamical sway of the giant planet. Despite this, there remains a large region in the inner parts of the HD\,83443\,c system that is dynamically stable on million year timescales, offering significant scope for the existence of additional, as yet undetected, planets. 

\subsection{Could additional planets lurk in the stable area of the system?}
\label{sec:additionalplanets}

\begin{figure}[hbt!]
\plotone{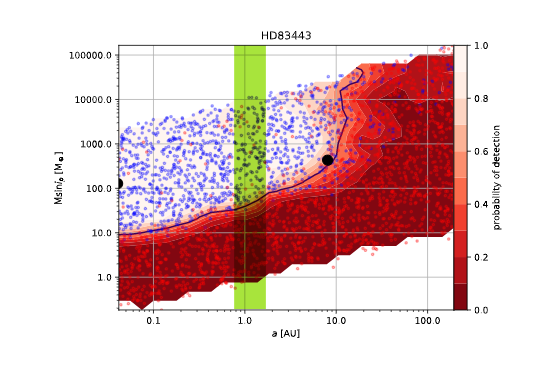}
  \caption{\texttt{RVSearch} results for detectability of additional planets in the HD\,83443 system. HD\,83443b and c are shown as large black points, and the injected test planets are shown as small blue points. The solid black line indicates a 50\% detection probability. HD\,83443c lies within the region of high detectability, and the radial velocity data can exclude additional planets larger than $\sim$ 10-40\me\ interior to 1\,au, the region shown to be dynamically stable.  The optimistic habitable zone is shaded in green.}
  \label{fig:recoveries}
\end{figure}

Having determined the regions in which additional planets are dynamically permitted, we ran injection-recovery simulations to set upper mass limits on the planets that can be ruled out by the radial velocity data \citep[e.g.][]{witt06, witt09, fulton2021}. For this analysis, we used \texttt{RVSearch} \citep{rosenthal21} to generate 3000 fictitious planets over a wide range of masses and orbital separations. The results are shown in Figure~\ref{fig:recoveries} as a detectability map. In the dynamically stable region interior to approximately 1\,au, the radial velocity data rule out the presence of planets more massive than 10-40\me, but smaller, potentially rocky planets may yet lurk undetected in this region (e.g. Section~\ref{habzone}).

\clearpage

\section{Discussion} \label{sec:Discussion}

\subsection{Searching for transits of the inner planet}

If HD\,83443 b transits, it is expected to have been observed by \tess\ given the short orbital period of the b planet ($\sim$3\,days) relative to the baseline of each continuous segment of \tess\ observations ($\sim$13\,day spacecraft orbit). Phase modulations may also be observed for HD\,83443 b given its short orbital period and the continuous orbital phase coverage by \tess. The expected semi-amplitudes of the modulations caused by Doppler boosting and ellipsoidal variations are expected to be $<$1\,ppm, which is below our detection limit for the \tess\ photometry. The semi-amplitude caused by atmospheric reflection and thermal emission of the planet is dependent on the planet's albedo and temperature, but could be as high as $\sim$150\,ppm for a perfectly reflective (1.0 geometric albedo) planet.

\begin{figure}[h!]
\plottwo{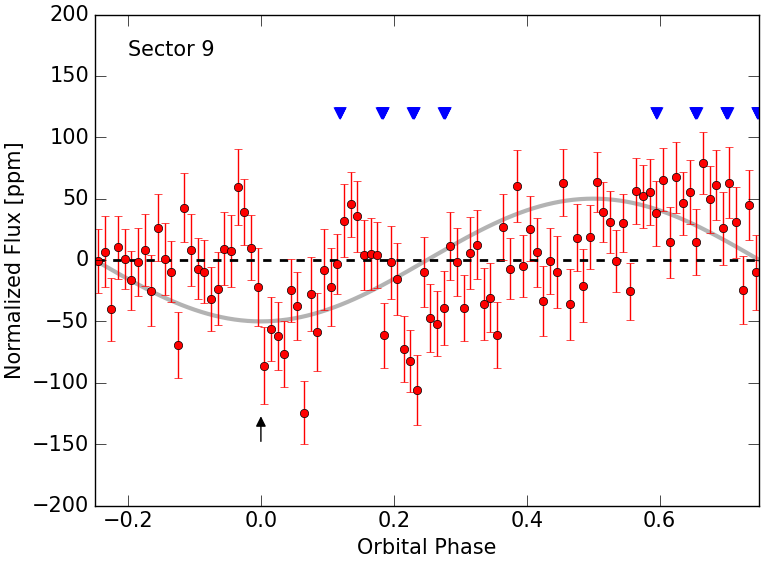}{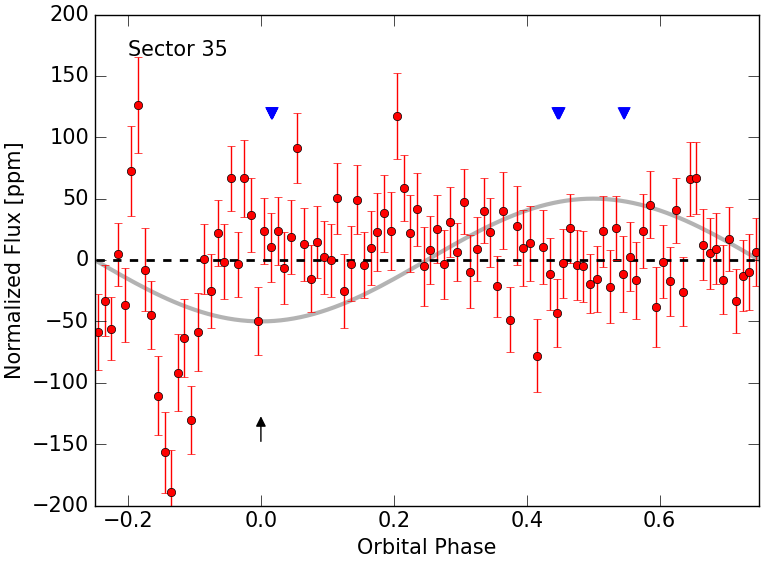}
\plottwo{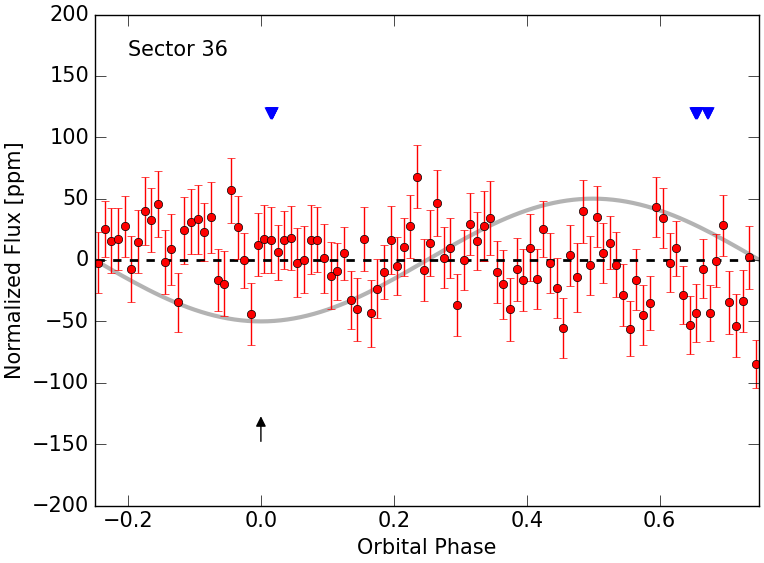}{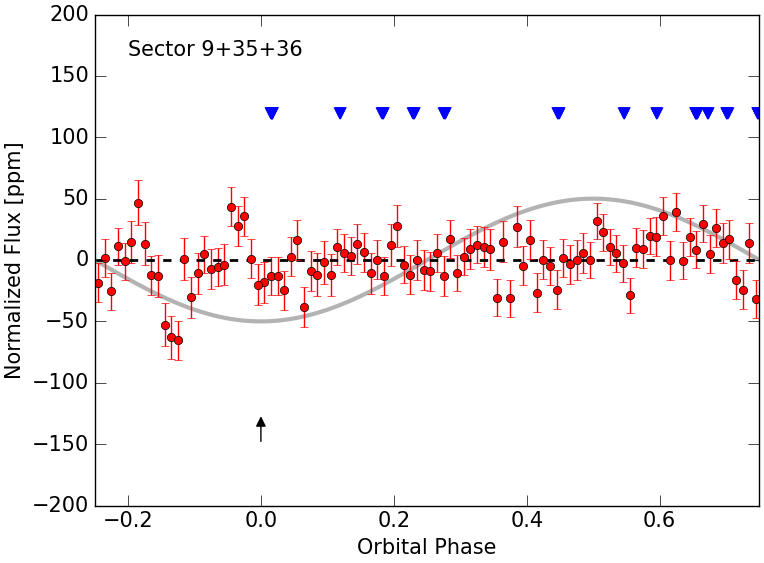}
\caption{The binned \tess\ PDC light curves phase folded on the orbital period of HD\,83443 b (2.985\,days) for Sector 9 (top left), Sector 35 (top right), Sector 36 (bottom left), and the concatenated light curve (bottom right). The red points show the binned photometry and their uncertainties show the standard deviation of the individual data points contained within each bin. The individual data points scatter beyond the shown flux limits and, thus, are not shown. The blue triangles indicate spacecraft thruster firing timings (i.e., momentum dumps). The black arrow indicates the expected time of transit at 0 phase and the gray curve shows an example of the expected phase modulations if the planet had 0.3 albedo. The observed modulations are inconsistent between \tess\ sectors and tend to align with momentum dump timings. Therefore, we conclude that we do not observe any evidence for transit events or phase modulations by HD\,83443 b.}
\label{fig:phasecurve}
\end{figure}

The \tess\ spacecraft obtained time-series photometry of HD\,83443 at 2-min cadence initially during Sector 9 observations (UT 2019 Feb 28 through 2019 Mar 25), and again nearly two years later during Sectors 35 and 36 (UT 2021 Feb 09 through 2021 Apr 02). The light curves are publicly available on the Mikulski Archive for Space Telescopes\footnote{\url{https://archive.stsci.edu/}} (MAST) and include the simple aperture photometry (SAP) and the pre-search data conditioning (PDC) light curves that were processed by the Science Processing Operations Center (SPOC) pipeline \citep{Jenkins16}. We searched for evidence of transit events and phase modulations caused by the inner planet in the \tess\ photometry using the orbital period and time of conjunction from the radial velocity analysis  ($T_c=2455000.3216^{+0.0068}_{-0.0076}$\,BJD). The PDC light curves from each individual \tess\ sector and the concatenated light curve were folded in phase to the period of the b planet such that the anticipated transit of HD\,83443 b would occur at 0 phase, and are shown in Figure~\ref{fig:phasecurve}. The PDC photometry exhibits modulations up to 100\,ppm, but the observed maxima and minima of the modulations may be consistent with the timings of thruster firings by the \tess\ spacecraft (blue triangles). The thrusters are used to periodically (every $\sim$2--3\,days) stabilize the spacecraft during observations (i.e., momentum dumps), but these events can cause systematic periodicities in the extracted light curves. Figure~\ref{fig:phasecurve} also shows an example of the anticipated phase curve that could be observed, with modulations of $\sim$50 ppm, if HD\,83443 b had a 0.3 albedo. However, hot Jupiters with observable phase curves typically exhibit albedo measurements that are nearly zero \citep[e.g.,][]{Wong2020, Wong2021, Kane2020}. Overall, the observed phase curve of HD\,83443 b differs between individual \tess\ sectors, has a shape that is inconsistent with either a transit event or possible emission/reflection modulations by the planet, and is consistent with being flat in the concatenated \tess\ light curve up to 20\,ppm. Therefore, we rule out transits and phase modulations of the inner planet, HD\,83443 b.


\subsection{Dynamical mass of HD\,83443c from Astrometry}
\label{sec:astrometry}


We have used \textit{Gaia} and \textit{Hipparcos} long-baseline proper motion anomalies from the eDR3 version of the \textit{Hipparcos}-\textit{Gaia} Catalog of Accelerations (HGCA; \citealp{Brandt2018_HGCA, Brandt2021_HGCA}) to break the sin\,$i$ degeneracy and measure the dynamical mass of HD\,83443c. HD\,83443 is not accelerating in the eDR3 version of the HGCA (its astrometric $\chi^2$ is only 0.15), indicating that a single-star solution fits the observed sky path extremely well. However, due to the excellent precision for HD\,83443 in \textit{Gaia} eDR3 (RUWE=0.939), HD\,83443c should have induced a detectable astrometric proper motion anomaly at nearly all masses above its radial velocity minimum mass. This means that although HD\,83443 is not an astrometric accelerator, we can exclude a vast range of masses above the radial velocity minimum mass and arrive at a strong dynamical mass constraint. This firmly places HD\,83443c into the planetary regime. 

We estimate the astrometric mass of HD\,83443c using the same methods as \citet{Brandt_2021_HR8799} and \citet{Dupuy_2021_51erib}.
We first compute an astrometric mass posterior. We draw five thousand trial orbits for HD\,83443c, adopting as priors all the posteriors on the fitted elements from the radial velocity analysis. We adopt a uniform prior on $\Omega$ and a geometric (sin\,$i$) prior on the inclination $i$, because these two elements are not constrained by the radial velocities. We use a Gaussian prior on the stellar mass of $1.00 \pm 0.03 \msun$, identical to the measurement by \citet{Delgado_Mena_2019}. 

\begin{figure}[hbt!]
\plotone{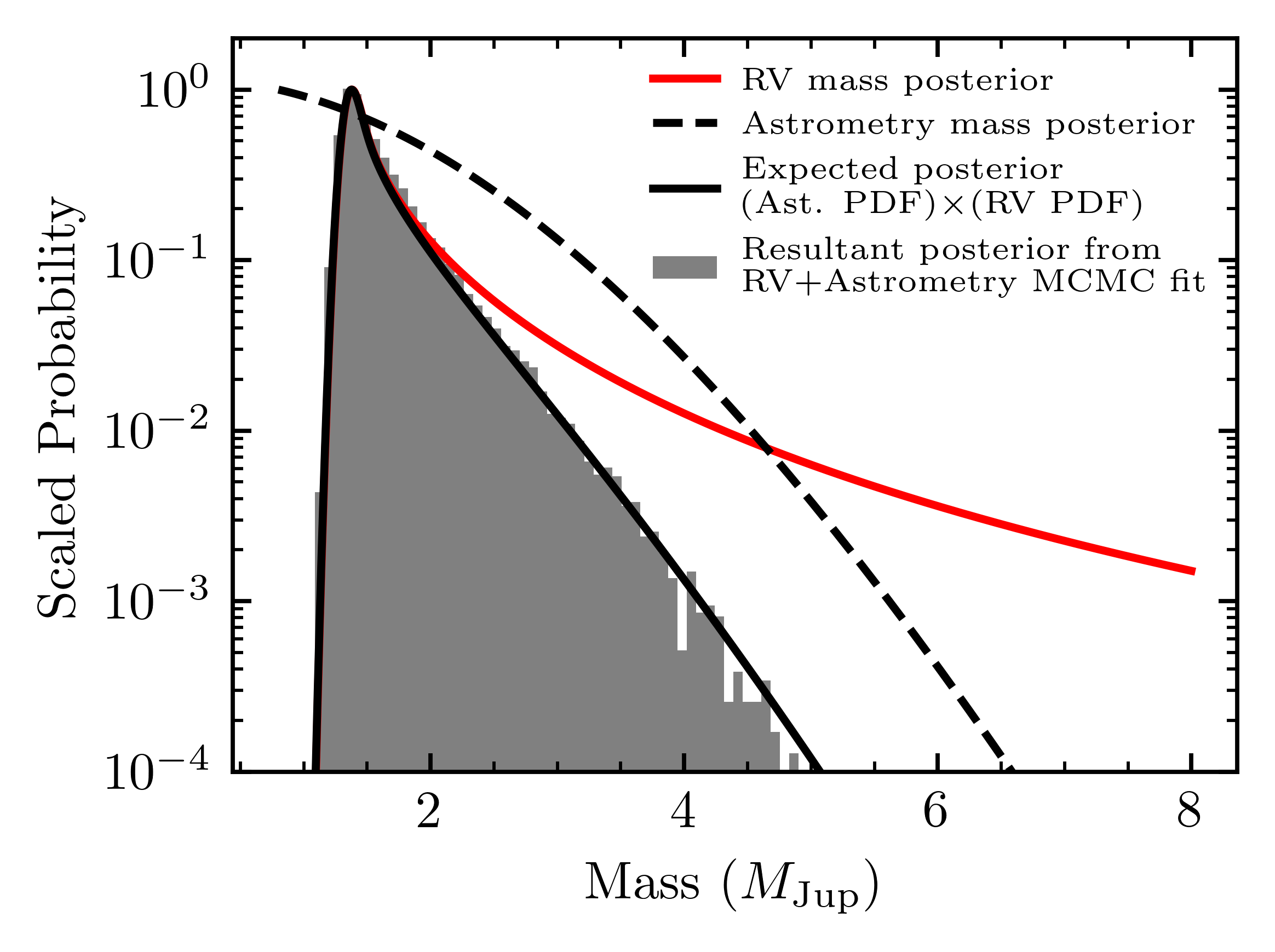}
\caption{The mass of HD\,83443c is the product of a lower limit set by radial velocities and a tight, astrometric upper limit constrained by its lack of acceleration between \textit{Hipparcos} and \textit{Gaia} eDR3. In red is the mass posterior from fitting radial velocities alone (i.e., the Gaussian m sin\,$i$ constraint convolved with the sin\,$i$ degeneracy). The black dashed line is the mass posterior implied by the lack of astrometric acceleration. The solid black line is the product of the red radial velocity posterior and the dashed-black astrometric posterior. The grey histogram is the resulting mass posterior from a joint orbital fit to the astrometry and radial velocities using \texttt{orvara}. The derived mass for HD\,83443c is $1.5^{+0.5}_{-0.2} M_{\rm Jup}$ with 1$\sigma$ confidence and $1.5^{+2.0}_{-0.3} M_{\rm Jup}$ with 3$\sigma$ confidence. }\label{fig:mass_posterior_astrometry}
\end{figure}

For each trial orbit, we solve for the best fit mass and error on that mass (see Section 2 of \citealp{Brandt_2021_HR8799}) using \texttt{htof} \citep{Brandt_2021_htof, Brandt_2021_htof_zenodo} version 1.0.1. We then add the posteriors from each of the 5000 orbital draws to arrive at the astrometric mass posterior for HD\,83443c. This mass posterior is shown by the dotted line in Figure \ref{fig:mass_posterior_astrometry}. The posterior is peaked at zero, an attribute owed to the excellent cross calibration of the HGCA and the fact that this source is a non-accelerator. The mass of HD\,83443c is constrained to be less than 3.5$M_{\rm Jup}$ with 99.7\% confidence from astrometry alone.

Figure \ref{fig:mass_posterior_astrometry} shows in red the posterior on the mass from the radial velocity fit. The long tail to high masses is due to the sin\,$i$ degeneracy, where high masses are geometrically disfavored. Multiplying the radial velocity mass posterior by the astrometry mass posterior results in the black curve of Figure \ref{fig:mass_posterior_astrometry}, which is the expected mass posterior that would result from a joint orbital fit to both the radial velocities and proper motion anomaly. Effectively, the final mass posterior is the astrometry mass upper limit (which excludes all masses above 3.5$M_{\rm Jup}$ with 3$\sigma$ confidence) cut off on the lower end by the radial velocity minimum mass (which allows only masses above $1 M_{\rm Jup}$ with 3$\sigma$ confidence). The black posterior is our final mass estimate for HD\,83443c: $1.5^{+0.5}_{-0.2} M_{\rm Jup}$ (1$\sigma$ confidence interval).

We confirm this mass estimate by performing a 3-body joint orbital fit of the radial velocities and astrometry using \texttt{orvara} \citep{TBrandt2021_orvara}. \texttt{orvara} employs MCMC with {\tt ptemcee} \citep{Foreman-Mackey_2013,Vousden_2016}. Absolute astrometry is processed and fit for the five astrometric parameters by \texttt{htof} at each MCMC step. We use a parallel-tempered MCMC with 20 temperatures; for each temperature we use 100 walkers with 1 million steps, thinned by a factor of 200. We use the standard, uninformative, priors on each of the orbital elements as discussed in \citet{Li_2021_RV}, except we adopt a uniform prior on planet mass instead of the standard 1/M prior. Convergence is assessed by the same criterion as in \citet{Li_2021_RV} and \citet{Brandt_2021_EDR3_masses}.

\begin{figure}[hbt!]
\plotone{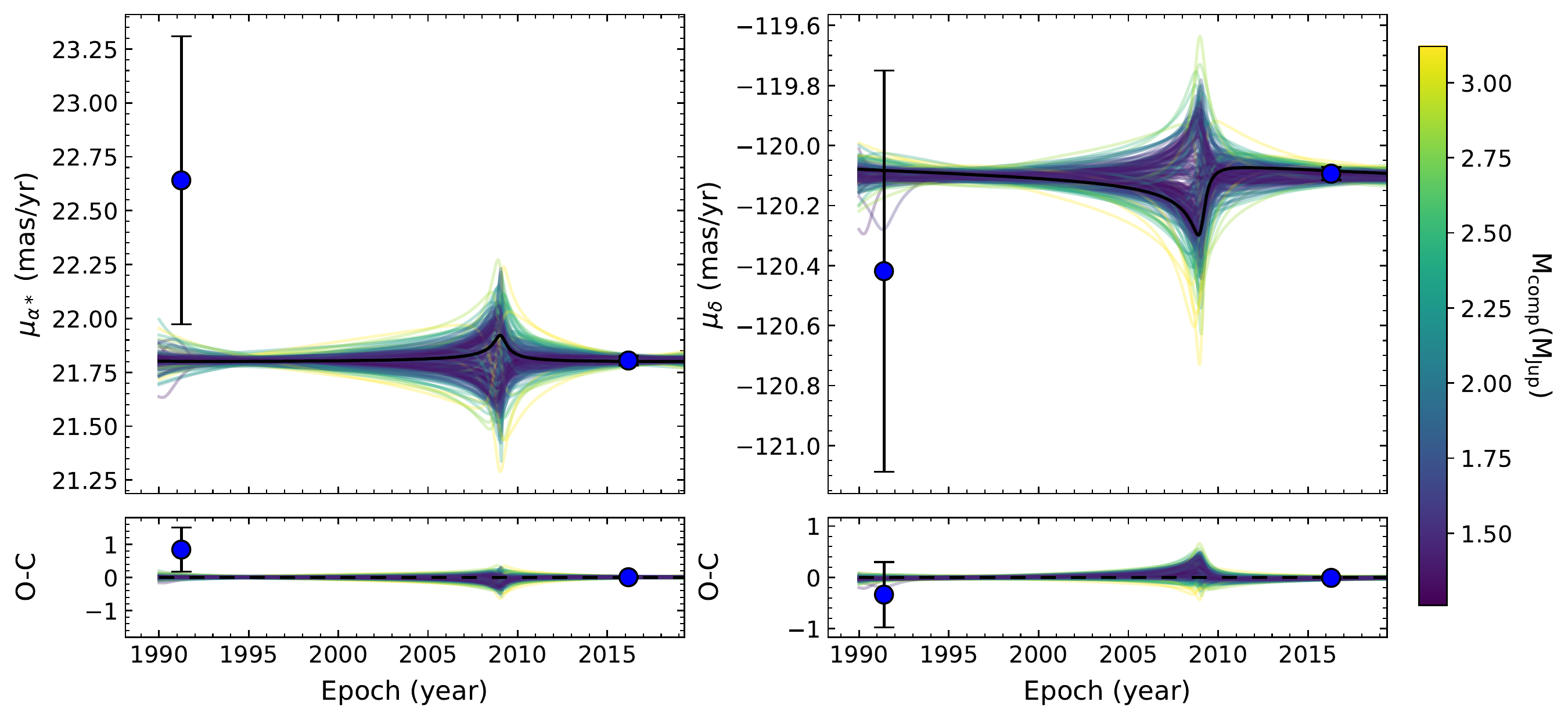}
\caption{Note to editor: this figure is new. Model proper motions compared to the calibrated \textit{Hipparcos} (point at 1991.25) and \textit{Gaia} EDR3 proper motions (point at 2016) from the HGCA. The best fit orbit is shown in black. A random sampling of other orbits from the MCMC chain are shown and are color coded by the mass of HD~83443~c. Note that the fit to the long-baseline proper motion between \textit{Gaia} and \textit{Hipparcos} is not shown because it is an integral constraint.}\label{fig:proper_motion_anomalies}
\end{figure}

The result of the joint MCMC fit is shown by the grey histogram in Figure \ref{fig:mass_posterior_astrometry}. The fits to the proper motions are displayed in Figure \ref{fig:proper_motion_anomalies}, where higher masses (yellow traces) result in proper motion accelerations that are further and further disfavored by the data. The MCMC mass posterior agrees nearly perfectly with the product of the radial velocity and astrometry mass posteriors. The derived orbital parameters are consistent with those in Table \ref{tab:MCMC_param}. The inclination posterior is constrained to be $90 \pm 33 \deg$, and the crucial extreme values of inclination ($<15\deg$ and $>165\deg$) are disfavored by factors of at least $e^{20}$ in likelihood. The mass, separation, and age of HD\,83443~c makes it a near-Jupiter analogue. 

This mass posterior can be improved over the coming decade with more data; but it is unlikely to improve significantly for the following reasons. First, because HD\,83443 is already in the optimal magnitude range for \textit{Gaia} ($G$ mag of 8). The astrometric precision ($\approx$0.01 mas in both right-ascension and declination) for HD\,83443 is projected to improve by only a small factor by mission-end \footnote{\url{https://www.cosmos.esa.int/web/gaia/sp-figure1}}. Improvements to the astrometric mass limit could be as great as a factor of two, but likely not greater. Second, the period is 22 years long, and so we need another $\sim$10 years of radial velocity to improve noticeably. And third, this mass measurement and the system's $\approx$5 Gyr age implies a magnitude so faint that we will not obtain direct imaging in thermal emission on this planet soon, even with the James Webb Space Telescope.

\subsection{Direct Imaging Prospects and the Habitable Zone} \label{habzone}

\begin{figure}[hbt!]
\plotone{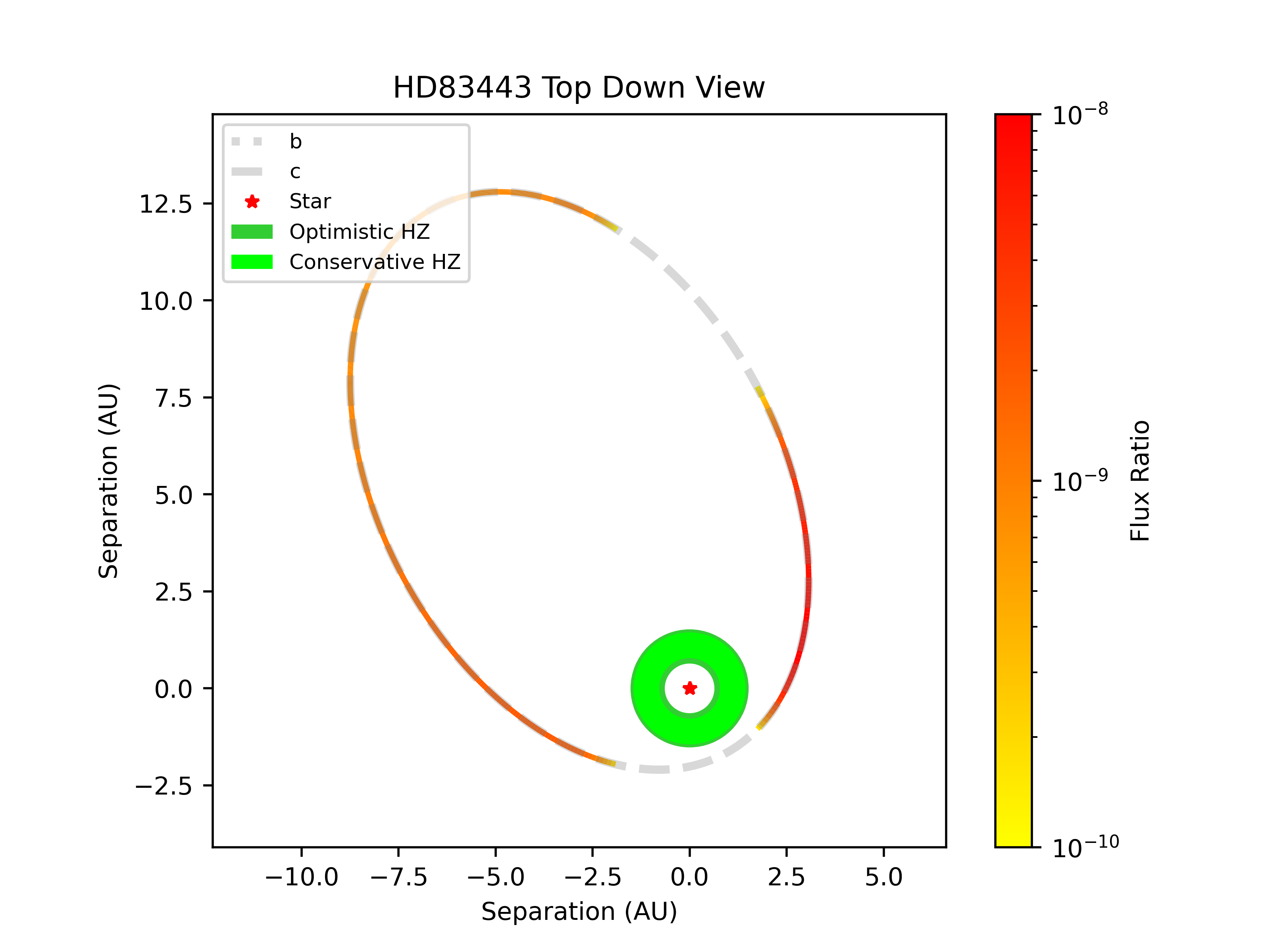}
\caption{Top down view of the HD\,83443 system with both conservative and optimistic HZ regions shown in light green and dark green, respectively. The orbit of the outer planet is color coded with the planet-to-star flux ratio if the brightness is above 1$\times$10$^{-10}$, assuming near edge-on inclination. Part of the orbit with low flux ratios not color coded is due to the planet passing through the inner working angle of \textit{HabEx} starshade. The inner b planet is not shown due to its proximity to the host star and the scale of the plot.}
\label{fig:DIContrast}
\end{figure}

An interesting aspect worth exploring for the newly discovered outer planet is its feasibility to be directly imaged by future space-based direct imaging missions thanks to the large angular separation and high eccentricity of its orbit \citep{kane2013, kane2018}. To calculate the brightness of the outer planet, we used the orbit visualization tool described by \citet{LiDI21} to carry out a first-order estimation of the planet-to-star flux ratio (i.e., without taking into account noise sources such as background stars and sky noise, exozodiacal dust, residual starlight, detector noise etc.). The flux ratio estimation assumed a Lambert sphere phase function, a planetary radius of $\sim$1.2~$R_J$ (estimated from the derived mass at edge-on inclination using a mass-radius relation by \citet{Chen2017}), and a geometric albedo of 0.5, consistent with previous estimates for Jupiter analogs \citep{cahoy10}. We further assume an edge-on inclination (90$^\circ$) from the astrometric result provided in Section~\ref{sec:astrometry}, and a HabEx Starshade mission concept configuration in the visible band (450-975~nm) \citep{habexreport}. To account for the uncertainty in the inclination derivation from astrometry, we calculated flux ratio variation of the outer planet for inclinations from perfectly edge-on to face-on cases. The planet could achieve flux ratios larger than $10^{-10}$, the required contrast ratio for HabEx \citep{habexreport}, outside the inner working angle (IWA) for all inclination cases. But due to the high eccentricity nature of the orbit and the location of the periapsis angle, the planet could have a maximum brightness with flux ratio around 9.5$\times$10$^{-9}$ near the edge of IWA for the 90$^\circ$ orbit. Figure~\ref{fig:DIContrast} shows the top down view of the outer planet's orbit. Flux ratio variation of the planet throughout its entire orbit for the edge-on case is color coded if the flux ratio is above $10^{-10}$.  Although the host star is a bit far ($\sim$ 41 pc) and dim ($V$ mag $\sim$ 8), the planet could potentially be imaged by future direct imaging missions when it approaches periapsis, which will happen in about 10 years, thanks to its high planet-to-star flux ratio.


We also use the stellar properties from \citet{2019AJ....158..138S}, shown in Table~\ref{tab:star}, to calculate the extent of the conservative and optimistic Habitable Zone (HZ) regions \citep{kasting1993,kane2012,kopparapu2013,kopparapu2014,kane2016}. The conservative and optimistic HZ lie in the range 0.86--1.54~AU and 0.68--1.62~AU, respectively, and are represented in Figure~\ref{fig:DIContrast} by the light and dark regions. The periastron passage of the outer planet passes close to the HZ, but does not enter the HZ region. As demonstrated by the dynamical results shown in Figure~\ref{stab} and discussed in Section~\ref{sec:additionalplanets}, this orbital architecture allows for the possibility of potential terrestrial planets within the HZ that are able to retain long-term stability. However, it is worth noting that such terrestrial planets may experience significant non-zero eccentricity variations due to the perturbations from the known outer planet, similar to the scenario of the HR~5183 system, described by \citet{wreckingball}.

\begin{figure}
 \begin{center}
 \begin{tabular}{cc}
  \includegraphics[scale=0.55]{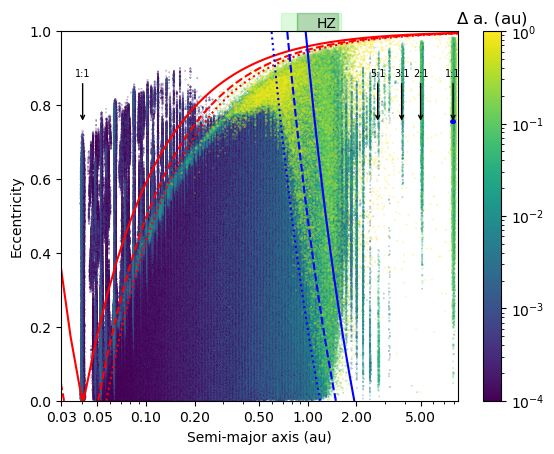} & \includegraphics[scale=0.55]{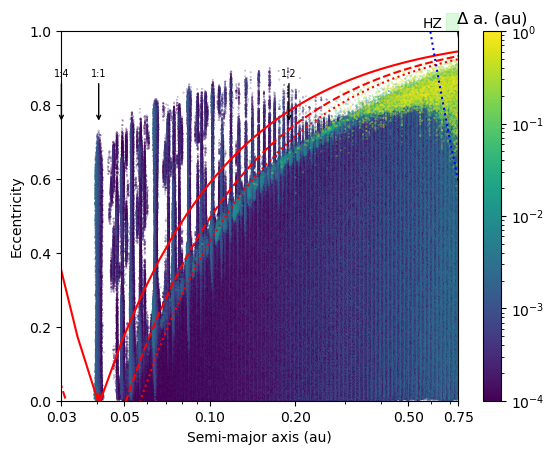} \\
  \includegraphics[scale=0.55]{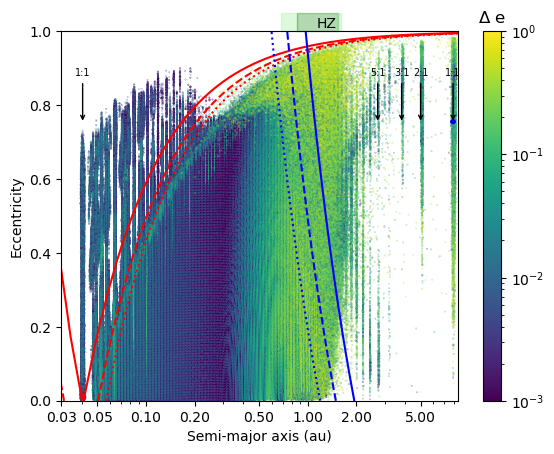} & \includegraphics[scale=0.55]{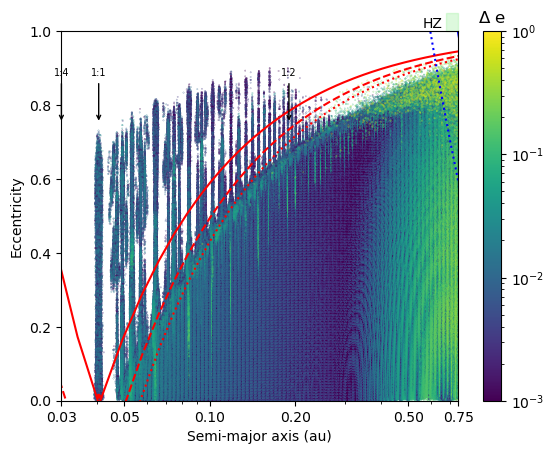} \\
  \end{tabular}
  \caption{The amount of stirring experienced by the stable test particles that survived the full duration of our simulations of the HD\,83443 system. Each individual dot plotted represents a test particle that survived until the end of our simulations. The top panels show, for each test particle, the difference between that particles initial and final semi-major axes (i.e. $\Delta a = |a_{init} - a_{final}|$), whilst the lower panels show the change, for each test particle, in orbital eccentricity (i.e. $\Delta e = |e_{init} - e_{final}$). The left panels show the full range of semi-major axes considered in this work, whilst the right panel zooms in on the inner 0.75~au of the system, to show in more detail the fine structure in that region. The solid blue line connects all $a-e$ locations with apocentre at the pericentre of the orbit of HD\,83443\,c, with the dashed and dotted blue lines marking orbits with apocentres one and three Hill radii interior to HD\,83443\,c's pericentre. Similarly, the solid red line connects all orbits with apocentre or pericentre at the orbit of HD\,83443\,b, such that all objects within the wedge described by the solid red line move on orbits that cross that of that planet. The red dashed and red dotted lines show orbits with pericentre three and five Hill radii beyond the orbit of HD\,83443\,b, respectively. Whilst orbits in the potential Habitable Zone around HD\,83443 are stable (in that test particles in that region can survive the duration of the simulations), it is apparent that those particles experience significant stirring, primarily from HD\,83443\,c, which would, at the very least, result in an planets in that region experiencing dramatic Milankovitch cycles, and may indicate that such orbits would prove unstable on longer timescales.}  \label{fig:Tim1}
 \end{center}
\end{figure}

In that light, it is interesting to further examine the results of the detailed dynamical simulations described in Section~\ref{sec:Dynamics}, to examine the degree to which the 'stable' particles (which survived for the full duration of the simulations) were stirred or excited by the influence of the two known planets. In Figure~\ref{fig:Tim1}, we therefore examine the degree to which the orbits of the surviving test particles are stirred by the end of our simulations. In the panels of that figure, we show the difference between the initial and final semi-major axes (top, $\Delta a$) and eccentricities (bottom, $\Delta e$) for all particles that survive to the end of the simulations. It is immediately clear that all test particles that are on even moderately eccentric orbits in the inner reaches of the system only survive if they are either (a) trapped in mean-motion resonance with HD\,83443\,b, or (b) move on orbits exterior to the red dotted line - orbits whose closest approach to the orbit of the inner planet occurs at a distance of more than 5 Hill radii exterior to that planet's orbit. 

Both the $\Delta a$ and $\Delta e$ plots show a marked excitation just beneath the red dotted line - showing that particles are being ejected from the inner part of the system on orbits with constant pericentre located just beyond that location. This is not a surprise - encounters with HD\,83443\,b act to excite the orbital eccentricity of test particles, which in turn modifies their orbital semi-major axis, since the encounters are happening at, or near, the pericentre of the particle's orbit. The result is that particles will random walk along that ejection line, with encounters increasing or decreasing the eccentricity and semi-major axis of the particle's orbit until either it encounters HD\,83443\,c, and is shifted off that line, decoupling it from the influence of HD\,83443\,b, or until it is ejected from the system entirely as a result of a kick from HD\,83443\,b. 

Such behaviour actually explains the population of test particles in the HZ of the HD\,83443 system that display marked excitation in both $a$ and $e$. Particles on highly eccentric orbits with $a \gtrapprox 0.6$ au can experience relatively close encounters with both HD\,83443\,b (at pericentre) and HD\,83443\,c (at apoapse). The result is that particles ejected to that regime by HD\,83443\,b can be moved onto more circular orbits, within the HZ, by encounters with HD\,83443\,c - along the lines of constant apocentre distance denoted by the blue lines. As such, particles from the inner areas of the system can be temporarily trapped in the HZ. The reverse behaviour is also clearly true - particles originating on orbits within 5 Hill radii of HD\,83443\,c can be excited, by that planet, until their pericentre falls within 5 Hill radii of HD\,83443\,b - which can then either help to eject those particles (moving them outward along the line of constant pericentre), or can drag them to smaller eccentricities along that line, decoupling them from the influence of HD\,83443\,c. 

The result is that the outer reaches of the HZ in the HD\,83443 system is a chaotic place, dynamically stirred by HD\,83443\,c to such a degree that it seems highly unlikely any planet mass objects could survive there on sufficiently stable orbits to be considered truly habitable. At low eccentricities, however, it is clear from Figure~\ref{fig:Tim1} that, at least in the inner part of the HZ, particles can survive on long timescales with only negligible stirring, and so, with the system's current architecture, one cannot rule out the presence of planets moving on stable orbits in the HZ of the system. Even in the inner reaches of the HZ, though, there is evidence of marked stirring in orbital eccentricity (with $\Delta e$ values of $\sim$ 0.1 or greater), which might suggest that, although planets in this region would be dynamically stable, they might nevertheless experience Milankovitch cycles of such amplitude as to render them effectively inhospitable. Should such planets eventually be discovered, simulations such as those described in \citet{Milankovic} and \citet{Milan2} will be vital to assess the severity of the climate variations that such behaviour would cause, and to help assess the degree to which those planets would be suitable target for follow-up observations.

\subsection{Origins of the HD\,83443 system}

The origin of hot Jupiters is still a topic of much discussion. As described in the Introduction, there are three main mechanisms that are invoked to explain the presence of hot Jupiters, namely:
\begin{itemize}
\item The inward migration of a giant planet through interaction with circumstellar disk; typically maintaining low eccentricity and inclination throughout.
\item One or more close encounters between two giant planets scattering those planets onto highly eccentric orbits, followed by tidal circularisation of the proto-hot Jupiter; typically results in low to moderate orbital inclinations for the hot Jupiter, and the presence of an outer planet on an eccentric orbit.
\item Kozai-Lidov perturbations on the orbit of the proto-hot Jupiter from a distant, inclined stellar companion, followed by tidal circularisation; typically results in strongly misaligned hot Jupiters.
\end{itemize}

\begin{figure}[hbt!]
  \plotone{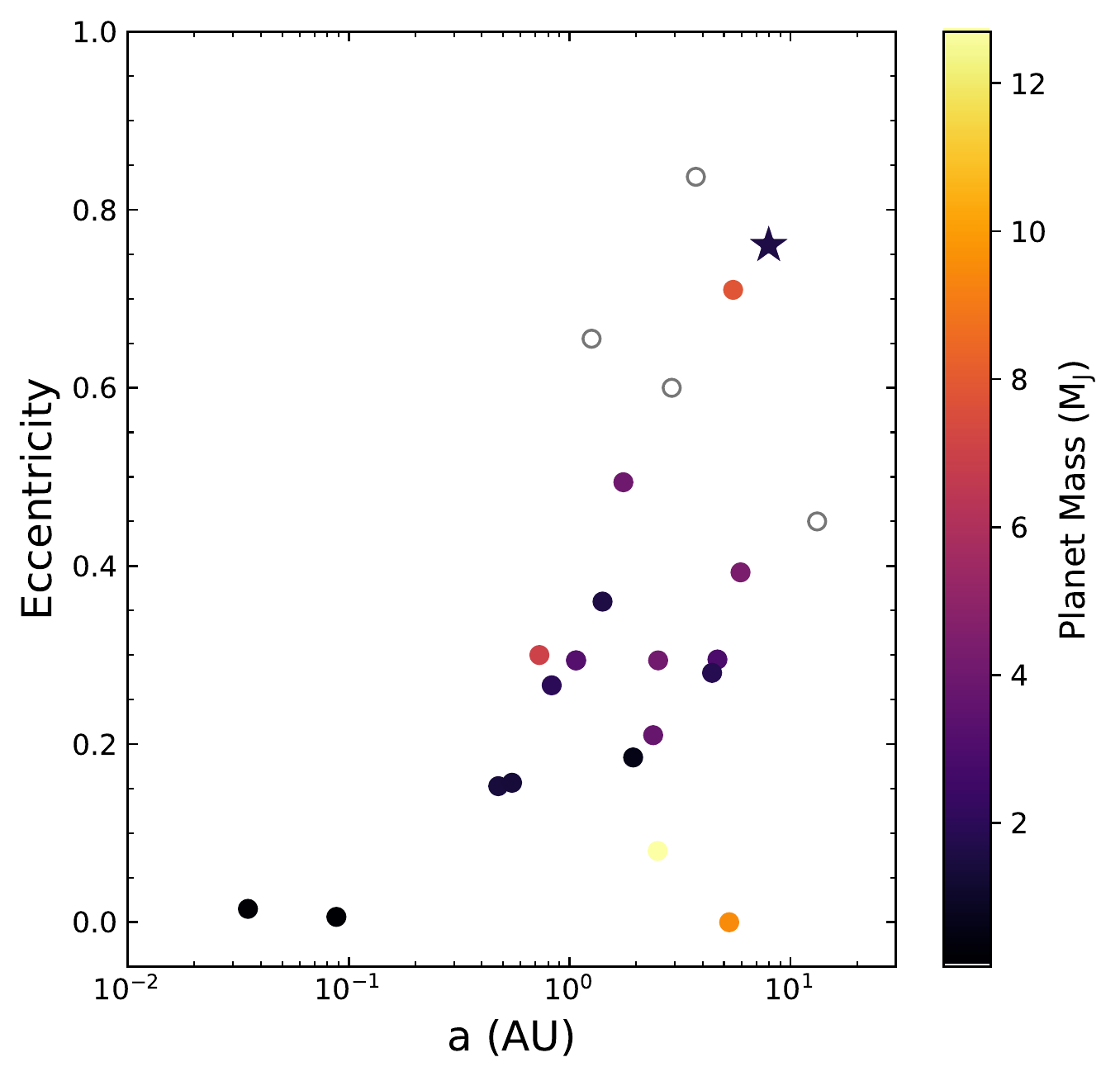}
  \caption{Eccentricity-semimajor axis distribution of exterior companions to Hot Jupiters. HD\,83443c is shown as a star, and is the most distant, most eccentric such planet yet found. Each planet is color-coded by its planetary mass value (or projected mass value if no radius measurement). The open circles represent known brown dwarf exterior companions. Planet data from NASA Exoplanet Archive, accessed 2021 Nov 15. }
  \label{fig:friends}
\end{figure}

In this context, the presence of HD\,83443\,c in its highly eccentric orbit is of particular interest, since it points to the possibility that HD\,83443\,b achieved its current hot Jupiter status as a result of chaotic encounters with that outer planet (i.e., the second mechanism noted above). As a consequence, the remaining planet is left in a short and highly eccentric orbit \citep[e.g.][]{1996Sci...274..954R, 1996Natur.384..619W, chatterjee08} and undergoes tidal circularisation \citep{nagasawa08, bonomo17, toi3362}.  
\citet{2011ApJ...735..109W} proposed a secular migration mechanism to explain the pile-up of hot Jupiters on $\sim$3-day orbits, the generally lower masses ($<$1\,\mj) of hot Jupiters, and the low frequency of additional planets in hot Jupiter systems within a few au (and noting that more distant eccentric companions were likely). The HD\,83443 system, with a 3-day hot Jupiter and an eccentric outer planet, could be considered a textbook example of the type of system produced by secular migration. It is worth noting that the analysis given in \cite{2011ApJ...735..109W} as well as \cite{chatterjee08} focused on systems initially containing three giant planets (where the third planet is ejected). The lack of further giant planets in the HD\,83443 system is consistent with such a history. Figure~\ref{fig:friends} shows the distribution of ``acquaintances of hot Jupiters'': systems containing a hot Jupiter and a distant exterior giant planet. Of particular relevance are those where the outer planet retains a high eccentricity. HD\,83443c is now the most distant, highest-eccentricity such planet. Only WASP-53c is more eccentric at $e=0.837$, although that object is a brown dwarf of at least 16.4\,\mj\ \citep{triaud17}. 

Many simulations and analysis have been done to reproduce the eccentricity distribution observed in the population of exoplanets, which includes the configuration of HD\,83443 \citep[e.g.][]{ford08, chatterjee08, 2010ApJ...711..772R, 2017A&A...598A..70S, 2019A&A...629L...7C, bowler20}. The Kozai-Lidov mechanism \citep{kozai, lidov} is often invoked to explain high-eccentricity planets \citep[e.g.][]{ecc1, witt07, blunt2019radial, venner21}. However, Kozai-Lidov oscillations, whereby orbital inclination is exchanged with eccentricity to produce high-eccentricity planets, require the influence of a binary stellar companion \citep{mustill17}. As shown in Section~\ref{sec:astrometry}, there is no evidence for a stellar-mass object bound to HD\,83443. We must instead look to an ancient and dramatic dynamical event to assign responsibility for the configuration of the HD\,83443 system. 

For systems in which many giant planets initially form, those planets perturb each other, which can result in eccentricity excitation and the possible ejection of planets by close dynamical encounters \citep[e.g.][]{1996Sci...274..954R, 1996Natur.384..619W, scat3, ford08}. \citet{2019A&A...629L...7C} showed that planet-planet scattering can result in final eccentricities even as high as 0.999. A scattering origin for the current orbits of the two planets in the HD\,83443 system would suggest that, despite the broad areas of stability revealed in the dynamical analysis described in Section~\ref{sec:Dynamics}, the presence of additional planets between the two described in this work is unlikely. Such planets would likely have been ejected or devoured through collisions during the period that the orbit of HD\,83443\,b was highly eccentric, particularly given that its orbital semi-major axis (and thus the location of all its associated mean-motion resonances) would have swept through the 'stable' region revealed by those simulations.

\citet{dawson13} showed that metal-rich stars hosted more hot Jupiters and more highly eccentric giant planets; the HD\,83443 system satisfies all of these characteristics. By comparison, the 14 Herculis system, with two giant planets \citep{14her, rosenthal21}, was recently fully characterised by \citet{bg21} with a combination of long-term radial velocity and astrometry. The outer planet was found to have a high eccentricity ($e=0.64^{+0.12}_{-0.13}$) and to be significantly misaligned with the inner planet. Much like for HD\,83443, the evidence strongly supports a scattering origin for the configuration of the 14~Her system. In the absence of Kozai-Lidov oscillations to drive eccentricity, it is highly probable that HD\,83443c arrived at its present location by a scattering event. It is interesting to note that scattering events would logically send the less-massive planet outward (as seems to be the case for 14~Her), yet for HD\,83443 the outer planet is by a factor of $\sim$3 the more massive of the two. This suggests that perhaps a third, even less massive ($<0.4$\mj) planet may have participated in the scattering and was ejected. 

\section{Conclusions}

We have combined more than 20 years of precise radial-velocity data from four instruments to detect a highly eccentric giant planet in a 22.6-year orbit around the nearby K dwarf HD\,83443.  The orbit of HD\,83443c is consistent with the $4\sigma$ positive linear trend noted by \citet{wright07}. We use the non-detection of astrometric acceleration to place a firm upper limit on the true mass of HD\,83443c, and with a joint fit of the radial velocities and astrometry we obtain a dynamical mass of 1.5$^{+0.5}_{-0.2}$\mj. This system was already known to host a Hot Jupiter, and is a rare example of a planetary system with a Hot Jupiter and a second giant planet.  The observed rarity of such objects is likely driven by the strong observational bias against the radial velocity detection of planets beyond $\sim$8-10\,au; imaging studies have derived occurrence rates of $\sim$50\% for such companions \citep{knutson14, bryan16}. HD\,83443c is reminiscent of the HR\,5183 system \citep{wreckingball}, in which a high-eccentricity giant planet exerts its dynamical influence to sweep clean a vast region. The inner $\sim$1\,au, including the HZ of this K0V star, remains stable to the presence of additional undetected planets. We have shown that the inner Hot Jupiter does not transit in \textit{TESS} photometry. The available astrometric data do not indicate any stellar-mass companions, further supporting the planetary interpretation of the radial velocity signal and suggesting emplacement in its current high-eccentricity orbit by planet-planet scattering rather than by the Kozai-Lidov mechanism. While HD\,83443c is beyond the grasp of planned imaging missions, it remains a fascinating example of a well-characterised system containing both a hot Jupiter and a highly eccentric giant planet.

\acknowledgments

{\textsc{Minerva}}-Australis is supported by Australian Research Council LIEF Grant LE160100001, Discovery Grants DP180100972 and DP220100365, Mount Cuba Astronomical Foundation, and institutional partners University of Southern Queensland, UNSW Sydney, MIT, Nanjing University, George Mason University, University of Louisville, University of California Riverside, University of Florida, and The University of Texas at Austin.

We respectfully acknowledge the traditional custodians of all lands throughout Australia, and recognise their continued cultural and spiritual connection to the land, waterways, cosmos, and community. We pay our deepest respects to all Elders, ancestors and descendants of the Giabal, Jarowair, and Kambuwal nations, upon whose lands the {\textsc{Minerva}}-Australis facility at Mt Kent is situated.

B.A. is supported by Australian Research Council Discovery Grant DP180100972. G.~M.~B. is supported by the National Science Foundation (NSF) Graduate Research Fellowship under grant no. 1650114.

T.F. acknowledges support from the University of California President's Postdoctoral Fellowship Program.

This paper includes data collected with the \tess\ mission, obtained from the MAST data archive at the Space Telescope Science Institute (STScI). Funding for the \tess\ mission is provided by the NASA Explorer Program. STScI is operated by the Association of Universities for Research in Astronomy, Inc., under NASA contract NAS 5–26555.

This publication makes use of The Data \& Analysis Center
for Exoplanets (DACE), which is a facility based at the
University of Geneva (CH) dedicated to extrasolar planets
data visualisation, exchange and analysis. DACE is a
platform of the Swiss National Centre of Competence in
Research (NCCR) PlanetS, federating the Swiss expertise
in Exoplanet research. The DACE platform is available at
\url{https://dace.unige.ch}.

%

\vspace{5mm}
\facilities{TESS, {\textsc{Minerva}}-Australis, AAT}


\software{AstroImageJ \citep{Collins:2017}, Exo-striker, \citep{2019ascl.soft06004T}, emcee \citep{Foreman-Mackey_2013}, htof \citep{Brandt_2021_htof, Brandt_2021_htof_zenodo}, orvara \citep{TBrandt2021_orvara}, ptemcee \citep{Foreman-Mackey_2013,Vousden_2016} } 







\clearpage





\clearpage
\begin{appendices}

\section{Cornerplots}\label{appendix}
\begin{figure*}[hbt!]
  \begin{center}
    \includegraphics[scale=0.14]{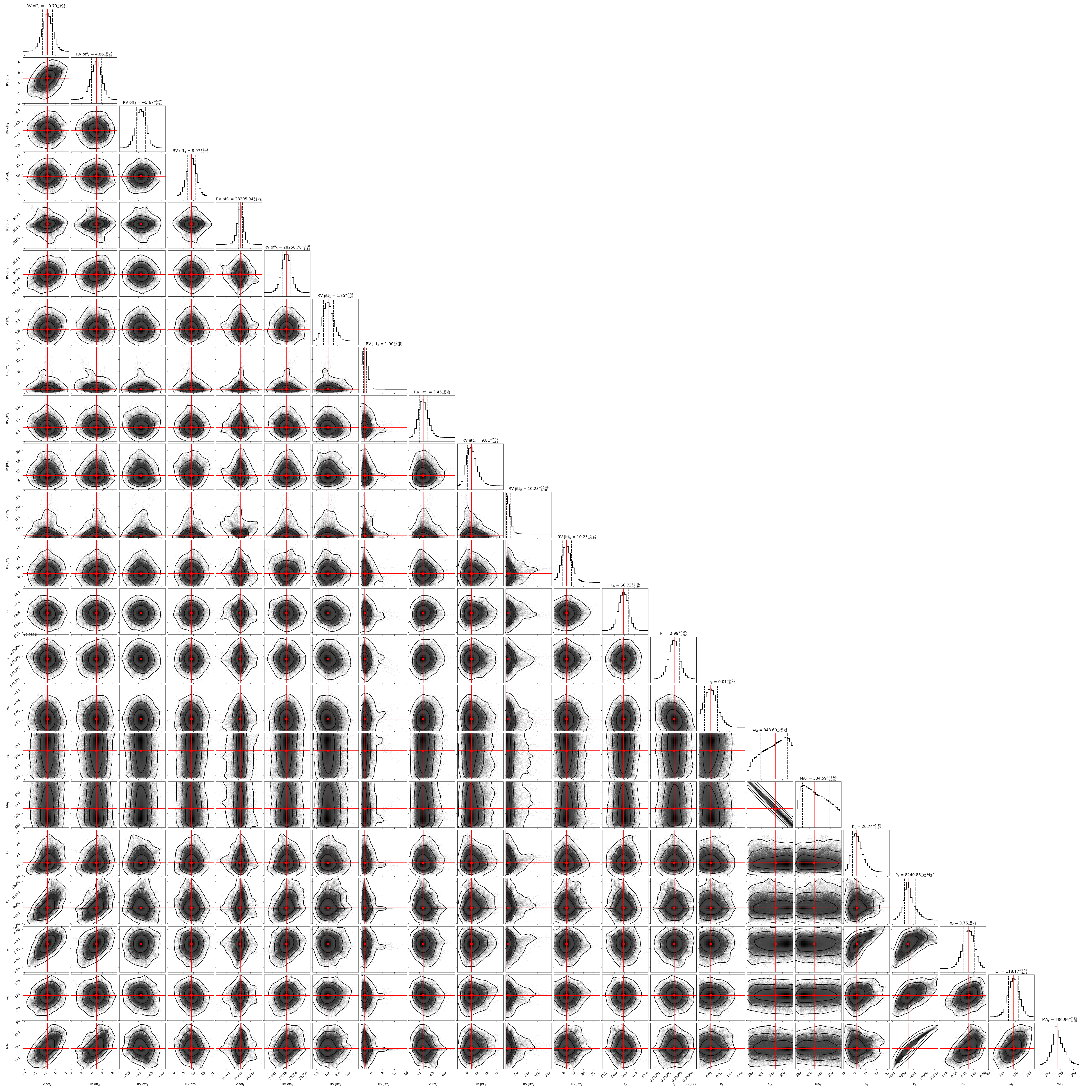}
    \caption{Corner plot - starting orbital period: 6000 days}
    \label{fig:cornerplot6000}
  \end{center}
\end{figure*}

\begin{figure*}[hbt!]
  \begin{center}
    \includegraphics[scale=0.14]{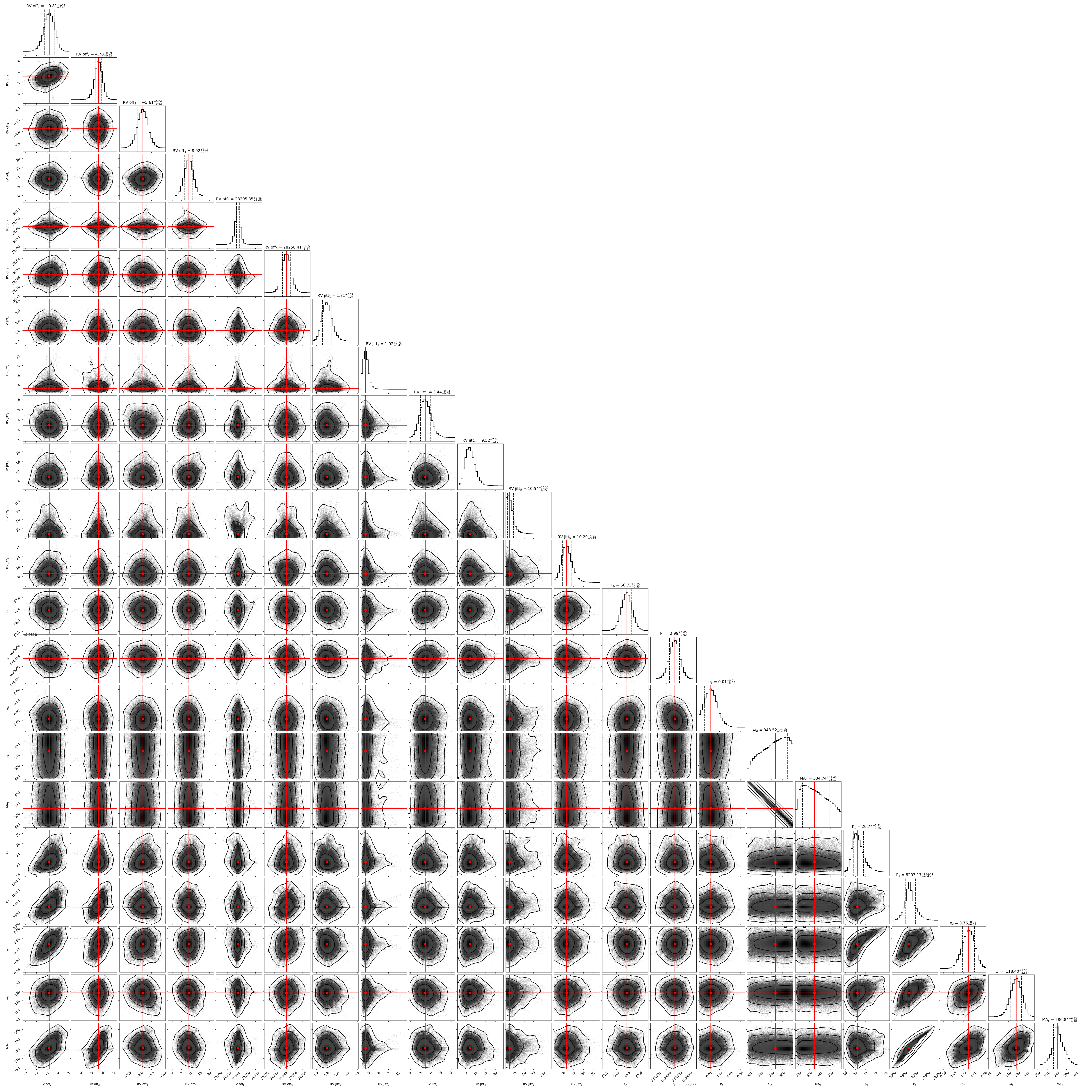}
    \caption{Corner plot - starting orbital period: 10000 days}
    \label{fig:cornerplot10000}
  \end{center}
\end{figure*}

\end{appendices}

\end{document}